\documentclass[a4paper,11pt]{article}

\usepackage[margin=1.14in]{geometry} 
\usepackage{tabularx}
\usepackage{multirow}
\usepackage{hyperref}
\usepackage{booktabs}

\usepackage{arydshln}
\usepackage{amsmath}
\usepackage{amsthm}
\usepackage{amssymb}
\usepackage{graphicx}
\usepackage{pdfpages}
\usepackage{float}

\usepackage{listings}
\usepackage{xcolor}

\definecolor{BrickRed}{rgb}{0.95,0.1,0.2}
\usepackage{hyperref}
\hypersetup{
	unicode,
	colorlinks,
	breaklinks,
	urlcolor=blue,
    linkcolor=black,
    citecolor=BrickRed,
	pdfproducer={LaTeX},
	pdfcreator={pdflatex}
}

\definecolor{codegreen}{rgb}{0,0.6,0}
\definecolor{codegray}{rgb}{0.95,0.95,0.95}
\definecolor{codepurple}{rgb}{0.0,0,0.8}
\definecolor{backcolour}{rgb}{0.95,0.75,0.92}
\definecolor{magenta}{rgb}{0.95,0.0,0.0}
\usepackage{courier}

\lstdefinestyle{mystyle}{
    backgroundcolor=\color{codegray},   
    commentstyle=\color{codegreen},
    keywordstyle=\color{codepurple},
    stringstyle=\color{magenta},
    basicstyle=\ttfamily\footnotesize,
    breakatwhitespace=false,         
    breaklines=true,                 
    captionpos=b,                    
    keepspaces=false,                 
    showspaces=false,                
    showstringspaces=false,
    showtabs=false,                  
    tabsize=2,
    lineskip=0.8ex
}
\usepackage{tabularx}
\usepackage{array}
\usepackage{booktabs}
\usepackage{wrapfig}
\usepackage[utf8]{inputenc}
\usepackage{lmodern}

\usepackage[T1]{fontenc} 
\usepackage{amsmath}

\newcommand{\be}{\begin{eqnarray}}
\newcommand{\ee}{\end{eqnarray}}
\def\be{\begin{equation}}
\def\ee{\end{equation}}
\def\bea{\begin{eqnarray}}
\def\eea{\end{eqnarray}}

\newcommand{\gsim}{\;\raisebox{-0.9ex}{$\textstyle\stackrel{\textstyle >}{\sim}$}\;}
\newcommand{\lsim}{\;\raisebox{-0.9ex}{$\textstyle\stackrel{\textstyle<}{\sim}$}\;}
\def\lsim{\raise0.3ex\hbox{$\;<$\kern-0.75em\raise-1.1ex\hbox{$\sim\;$}}}
\def\gsim{\raise0.3ex\hbox{$\;>$\kern-0.75em\raise-1.1ex\hbox{$\sim\;$}}}

\usepackage{graphics}

\usepackage{epsfig}
\usepackage{slashed}
\usepackage[utf8]{inputenc}
\usepackage{multirow}
\usepackage{dcolumn}

\usepackage[sort&compress,numbers,square]{natbib}

\theoremstyle{plain}

\theoremstyle{definition}

\usepackage{graphicx, color}

\title{\texttt{DLScanner}: A parameter space scanner package assisted by deep learning methods}
\vspace{8mm}
\author{\Large{A. Hammad$^{a}$$\thanks{\href{mailto:hamed@post.kek.jp}{hamed@post.kek.jp}}$ \ and Raymundo Ramos$^{b}$$\thanks{\href{mailto:raramos@kias.re.kr}{raramos@kias.re.kr}}$}}
\date{
{}
$^a$ Theory Center, IPNS, KEK,  1-1 Oho, Tsukuba, Ibaraki 305-0801, Japan.\\
$^b$Quantum Universe Center, Korea Institute for Advanced Study, Seoul 02455, Korea}

\begin{document}
	\maketitle
	\vspace{4mm}
	\begin{abstract}
 \normalsize{
In this paper, we introduce  a scanner package enhanced by deep learning (DL) techniques. The proposed package addresses two significant challenges associated with previously developed DL-based methods: slow convergence in high-dimensional scans and the limited generalization of the DL network when mapping random points to the target space.
To tackle the first issue, we utilize a similarity learning network that maps sampled points into a representation space. In this space, in-target points are grouped together while out-target points are effectively pushed apart. This approach enhances the scan convergence by refining the representation of sampled points.
The second challenge is mitigated by integrating a dynamic sampling strategy. Specifically, we employ a VEGAS mapping to adaptively suggest new points for the DL network while also improving the mapping when more points are collected.
Our proposed framework demonstrates substantial gains in both performance and efficiency compared to other scanning methods.
 }
\end{abstract}
\newpage
\noindent\rule{\textwidth}{1pt}
\tableofcontents
\noindent\rule{\textwidth}{0.2pt}
\maketitle \flushbottom
\vspace{4mm}

\section{Introduction}
\label{sec:intro}

The exploration of signatures beyond the standard model (BSM) at the Large Hadron Collider (LHC) continues very actively
after finally observing the Higgs boson in 2012.
These searches tend to be guided by our knowledge from leading BSM theories
pointing solutions to current missing pieces in the standard model.
The amount of analysis carried on the data obtained by the LHC
has resulted in tighter constraints on BSM theories,
although compelling signals of new physics have eluded all observations.
As a result, identifying viable parameter spaces within BSM theories has become increasingly challenging, often requiring comprehensive scans of their parameter spaces.

Several scanning approaches have been devised in the past. The most straightforward methods are grid and random sampling. While these approaches can explore the full parameter space, they suffer from slow convergence to target points, those that satisfy experimental and theoretical constraints. Adaptive sampling methods, in contrast, leverage likelihood-based approaches to focus on valid points by maximizing the likelihood function during each iteration of the scan. Well known examples include Markov Chain Monte Carlo (MCMC)~\cite{Speagle:2019ffr,wang2022effective,robert2011short,Berg:2004fd} and MultiNest~\cite{Feroz:2008xx,Feroz:2013hea}. Recently, a variety of publicly available scanning tools based on adaptive sampling methods have been released, such as Fittino~\cite{Bechtle:2004pc}, GAMBIT~\cite{Martinez:2017lzg}, BSMArt~\cite{Goodsell:2023iac}, EasyScan-HEP~\cite{Shang:2023gfy}, and others~\cite{Lewis:2002ah,RuizdeAustri:2006iwb,Allanach:2007qj,Strege:2014ija,Han:2016gvr,Bagnaschi:2017tru,Brinckmann:2018cvx,Diaz:2024yfu,Diaz:2024sxg}. These tools are specifically designed to use the results of likelihood calculations to infer distributions of model parameters and provide samples consistent with these distributions. 
While these methods are highly effective, they can encounter challenges when dealing with particularly complex subspaces or regions with problematic features. Such difficulties may result in excessive evaluations of the likelihood function or poorly sampled areas~\cite{Goodsell:2022beo}. Moreover, these methods often require considerable time to converge to the desired regions of parameter space~\cite{Ren:2017ymm,Hammad:2022wpq}.

Recently, deep learning (DL) methods have been proposed to address these challenges~\cite{graff2012bambi,Ren:2017ymm,Staub:2019xhl,Caron:2019xkx,Goodsell:2022beo,Hammad:2022wpq}. These approaches involve iterative scans, where a DL network is progressively trained on an expanding dataset of accumulated points. As the dataset grows, the trained network becomes increasingly accurate at predicting new points likely to lie within the target region. 
The process for identifying valid points depends on the architecture of the DL network, namely, a regressor or a classifier. In the case of a DL regressor, a likelihood function is employed to measure how closely the points predicted by the regressor to the target region. Conversely, a DL classifier relies on an oracle function to determine whether suggested points are ``valid'' (within the desired region) or ``invalid'' (outside the desired region). This is typically achieved by assigning binary labels, with valid points labeled as 1 and invalid points as 0. Moreover, DL regressor is designed to map each point in the sampled space to the target space while classifier is designed to learn the decision boundaries of the target space. Both methods bring improvements over adaptive sampling methods, e.g., MCMC and MultiNest, as the DL network can effectively learn a high dimensional nonlinear mapping between the  sampling and target space.  Nonetheless, DL-based methods face significant unresolved challenges, two of which are: slow convergence in high-dimensional parameter spaces and limited capacity of the DL network for generalization.

In this paper, we introduce a novel DL-based approach that combines a similarity learning (SL) network~\cite{Dillon:2021gag,Esmail:2023axd,Dillon:2023zac,Hammad:2024dsn,Hallin:2024gmt} with VEGAS mapping~\cite{Lepage:1977sw,Lepage:1980dq,Lepage:2020tgj}. Integrating the SL network into the scanning loop improves convergence when sampling from high-dimensional spaces. This is achieved as the SL network maps the sampled points into a representation space where valid points are grouped closely together, while invalid points are pushed farther apart. This is achieved by minimizing a contrastive loss function, built with a distance-based measure.
During training, the representation space is structured as a hypersphere, where minimizing the contrastive loss increases the Euclidean distance between valid and invalid points on the sphere. Once the distribution of sampled points is well-structured in this representation space, the network can predict new valid points in the target region.
A key advantage of the SL network is its two-step mapping process. Unlike previous DL approaches that directly map sampled points to the target space, the SL network first maps the points to a fixed-dimensional representation space and then maps them from this intermediate space to the target space. As a result, the SL network is primarily influenced by the dimensionality of the representation space, not the higher dimensionality of the sampled space.

Even with advanced DL networks like the SL network, the performance of predicting valid points from randomly sampled data remains suboptimal. To address this, a VEGAS map can be trained on accumulated points to generate new samples to suggest to the DL network. The VEGAS map could take care of generating samples concentrated on valid regions, accelerating convergence. Incorporating a VEGAS map into the scanning loop mitigates two key challenges: slow convergence during the initial iterations (when the number of valid points is limited) and suboptimal generalization caused by network structure or poorly optimized hyperparameters.

The proposed DL approaches are integrated into a Python-based package that we named \texttt{DLScanner},
aiming to be an user-friendly scanning tool. \texttt{DLScanner} includes both DL regressors and classifiers, which are enhanced by adaptive VEGAS maps. While \texttt{DLScanner} is developed as a versatile and generic scanning package, it also comes pre-configured with default modules for SPheno and \texttt{micrOMEGAs}.

This paper is organized as follows:
In Sec.~\ref{sec:terminology} we introduce the terminology of DL-based scanning methods and discusses the role of VEGAS in the iterative process.
In Sec.~\ref{sec:installation} we provide guidance on installation and how to start using the package.
In Sec.~\ref{sec:methods} we explain the sampling methods employed in the package.
In Sec.~\ref{sec:scanning} we explain the scanning modules integrated into \texttt{DLScanner}.
In Sec.~\ref{sec:mssm} we present results from a scanning example applied to the parameter space of the minimal supersymmetric standard model (MSSM).
Finally, we conclude in Sec.~\ref{sec:conclusion}.
We also include, in an Appendix, a small code example of the generic scanner.

\section{Terminology of the parameter space exploration}
\label{sec:terminology}                                       
The procedure begins by generating an initial set of random values, $K_0$, for the parameters under investigation, along with corresponding calculations of observables, $Y(K_0)$. This preliminary dataset of parameters and calculated results serves as the first training dataset for the DL network. The DL model purpose is to guide future iterations by identifying parameter regions that merit more detailed examination.
Following this initial training phase, an iterative process is conducted follows:

\begin{enumerate}
    \item \textbf{Zeroth step:} Accumulation of valid points from prior random scan and training of the DL network on these points. 
    \item \textbf{Prediction:} The DL network predicts the outcomes, $\hat{Y}$, for a large set of random points, $L$.
    \item \textbf{Selection:} A smaller subset of parameters, $K\subset L$, is chosen based on specific criteria aligned with the analysis goals, resulting in a set $K$ with network prediction $\hat{Y}(K)$.
    \item \textbf{Refinement}: The chosen parameters, $K$, are evaluated through the HEP package yielding the true values $Y(K)$.
    \item \textbf{Network update:} The true points are accumulated and the DL network is retrained using the accumulated dataset.
    \item \textbf{Iteration:} The refined network is used to generate new predictions for an expanded set of parameters, as in Step 2.
        This cycle is then repeated.
\end{enumerate} 

\begin{figure}[!h]
    \centering
    \includegraphics[width=\linewidth]{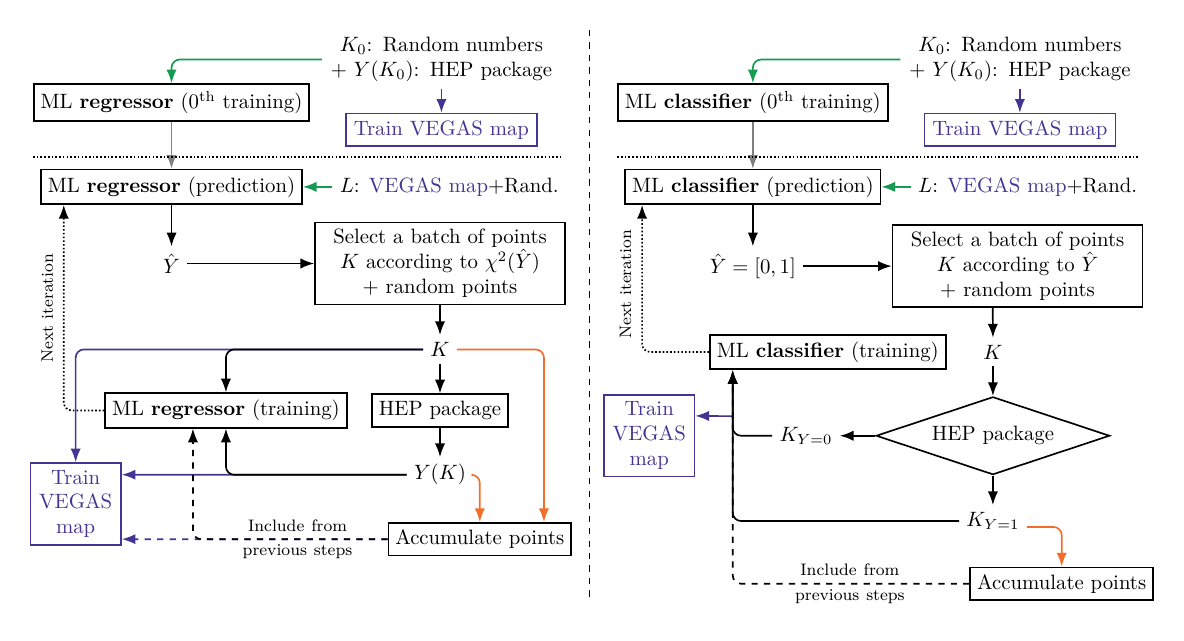}
    \caption{%
        \label{fig:chart}%
        Charts for the iterative processes used for the regressor (left)
        and the classifier (right).
        Black arrows indicate the main predict-train loop,
        green arrows indicate places where random input is required,
        orange arrows mark the parts where we collect the output dataset
        and blue arrows highlight where VEGAS mapping is trained and
        used to suggest new points.
    }
\end{figure}

In later sections we give details of specific implementations of this method. Given the high computational expense of calculating the observables, training a DL network to approximate results is worthwhile, enabling a more efficient sampling.
The sampled $K$ values and their corresponding output represent the evolution of exploring the parameter space. For each training in each iteration, new collected points are used, although accumulated points may also be selectively included. To manage computational load, training may use the full dataset periodically, based on a fixed rule. The type of incorporated DL networks determines the scanning results. The different processes are represented in Fig.~\ref{fig:chart} according to the approach. Two main DL approaches are used:

\begin{itemize}
    \item \textbf{Regression:} This approach aims to precisely maps each point in the sampled space to the target space. Points outside a target regions are rejected and the network is trained only on accumulated points in the target region.  The network needs to be trained on a large number of points to be able to find a encode an accurate mapping between parameters and their results. Understandably, this requires a large number of iterations. This approach can be based on a likelihood function to measure how close the sampled points are to the best fit value of the target with continuous output ranging from 0 to 1.
        In this approach, points can be selected for accumulation and refinement above some likelihood value.
    \item \textbf{Classification:} This approach trains an DL classifier to predict whether a point falls within a specified region of interest, such as being theoretically valid or satisfying experimental constraints. Here, a binary HEP calculation provides outputs, 0 or 1, for invalid and valid points, respectively. The DL classifier outputs predictions, $\hat{Y}$, between 0 and 1, indicating the probability of the sampled points to be in-target. Points can be selected based on high confidence values or areas of uncertainty, with $\hat{Y}> 0.5$ requiring refinement to determine the correct binary labels.
    Classification enables the creation of separate datasets for in-region and out-region points, simplifying training for boundary determination.
\end{itemize}

By systematically incorporating feedback from actual calculations, the DL model improves at identifying key areas.
For classification networks, different methods can be utilized. In this work we focus on multi-layer perceptron (MLP) and SL networks.  Using an MLP  network and supervised SL would offer different advantages, particularly in how they guide the iterative exploration of parameter space. 

An MLP network, acting as a  classifier, focuses primarily on learning the direct relationship between input parameters and target outputs, such as observables or region classifications. In each iteration of the loop, the MLP is retrained with updated data to improve predictions for new points and refine  classification boundaries. This approach is relatively straightforward and efficient for tasks with clear input-output relationships, as it can approximate the values of observables or determine if points lie within regions of interest. However, an MLP might struggle to generalize,
overfitting in densely sampled areas and underperforming in areas that remain less explored. Moreover, scanning over high dimensional space, we may need a more complex structure with higher capacity for abstraction.

On the other hand, supervised SL  refines the model by encouraging it to make a clear distinction between points, especially around decision boundaries.
In supervised SL, the model is trained to cluster representations of points within the same class closer together in representation space,
while keeping different classes far apart.
This approach leverages similarities and differences between points, making it particularly effective when separating decision boundaries.
Moreover, supervised SL improves the network ability to generalize by identifying fine boundaries within the parameter space, being useful when regions are complex or high dimensional. Additionally, this method aids uncertainty-based point selection, as it highlights points near decision boundaries where model uncertainty tends to be higher.
This is the main functional difference with MLP:While the MLP approximates the observables or regions directly, supervised contrastive learning enhances boundary delineation by focusing on separating similar and dissimilar points.
The choice between these methods in the scanning loop depends on the complexity of the decision boundaries within the parameter space and the importance of well-separated regions for the analysis.

In all considered approaches, it is possible for the model to overfit to the training set, leading to inaccurate predictions for unseen points. In our approach, we mitigate this issue by iteratively comparing the model predictions with the results from the true calculations and correcting regions where the model provides inaccurate guidance.
For the regressor, training with points in regions of high importance can result in large errors for predictions in regions of low importance. The goal of the outlined process is to minimize such errors while ensuring that no relevant regions are overlooked.  
The classifier operates differently, as it determines whether a given point belongs to the target region. Here, the trained model provides a measure of confidence regarding whether a new point lies inside or outside the target region. To ensure reliability, we require the model to achieve sufficient confidence in its predictions for both types of regions. This iterative refinement ensures robust performance across the parameter space.

\subsection{The role of VEGAS}

The VEGAS algorithm~\cite{Lepage:2020tgj}
is a well known method for adaptive multidimensional Monte Carlo integration.
One of the particulars of this algorithm
is the creation of an adaptive map
that changes to integration variables
that flatten the peaks of the integrand.
While we do not deal with integration variables,
we can take advantage of this mapping
to aid in the generation of points
in multidimensional spaces,
concentrating on regions of higher importance.
One of the disadvantages of problems that consider many dimensions
is the reduction in efficiency of points generated in-target
as number of dimensions grow.
By training a VEGAS map using points we know to be in-target (or close)
we can generate a large set of random points
with a more dense in-target distribution.
The network will still take care of deciding which of these
points have a larger probability of being in target.
The explicit places where the VEGAS map is used in our processes
are indicated in~\ref{fig:chart}.

Here we repeat a few details about this map.
For the full description the interested reader can check Ref.~\cite{Lepage:2020tgj}.
The VEGAS map is created by dividing coordinates in intervals.
Assume that we have a coordinate $x$ in the range $[a, b]$ that we want to divide into $n$ intervals:
\begin{align}
    x_0 & = a \nonumber\\
    x_1 & = x_0 + \Delta x_0 \nonumber\\
        & \ldots \nonumber\\
    x_n & = x_{n - 1} + \Delta x_{n-1} = b \nonumber
\end{align}
where $\Delta x_j$ are the widths of the intervals.
Now take a variable $y\in [0, 1]$, for which
\begin{equation}
    x(y) = x_{j(y)} + \Delta x_{j(y)} \delta (y)
\end{equation}
where
\begin{align}
    j(y) & \equiv \mathrm{floor} (y n)\\
    \label{eq:mapdeltay}
    \delta(y) & \equiv y n - j(y)\,.
\end{align}
In this case $y$ would be the transformed variable.
From Eq.~\eqref{eq:mapdeltay} it is easy to see that the intervals of varying width $\Delta x_j$ in $x$-space
become intervals of constant $1/n$ width in $y$-space.
This transformation has a Jacobian given by discrete values that depend on the interval widths $\Delta x_j$
\begin{equation}
    J(y) = J_{j(y)} \equiv n \Delta x_{j(y)}\,.
\end{equation}

The VEGAS module, available for python,
allows the creation of maps
from collected samples using the \texttt{adapt\_to\_samples} function.
We include submodule based on this
to define a function \texttt{vegas\_map\_samples}
that trains a map and return a function that uses it
to create a large sample.
The arguments and output of the function are as follows:
\begin{itemize}
    \item \texttt{DLScanner.utilities.vegas.\textbf{vegas\_map\_samples}}
    \begin{itemize}
        \item
            \texttt{xtrain}: array-like, coordinates of the sample.
        \item
            \texttt{ftrain}: array-like, output of the sampled function for \texttt{xtrain}.
        \item
            \texttt{limits}: array-like, boundaries of the coordinates being sampled.
        \item
            \texttt{ninc=100}: integer, optional, number of increments in \texttt{vegas} map.
        \item
            \texttt{nitn=5}: integer, optional, number of iterations using in adaptation.
        \item
            \texttt{alpha=1.0}: float, optional, Damping parameter for adaptation.
        \item
            \texttt{nproc=1}: integer, optional, number of processes/processors to use.
        \item On return: Callable function that takes an integer and outputs that many samples.
    \end{itemize}
\end{itemize}

For more information on the arguments \texttt{ninc}, \texttt{nitn}, \texttt{alpha}, and \texttt{nproc}
it is better to check the documentation of the \texttt{vegas} module~\cite{peter_lepage_2024_12687656}.
Internally, \texttt{vegas\_map\_samples} takes care of normalizing the ranges of \texttt{xtrain}
to the range $[0,1]$ to avoid problems with parameters with very large ranges.
Typically,
in our case we use this function with \texttt{xtrain} and \texttt{ftrain}
from the accumulated points during the iterative process.
Training a VEGAS map is fast so we recreate the map in every iteration.
This also allows to adjust the setup of the map
according to accumulated number of samples
and avoids bias from previous trainings that employed less samples.

\section{Installation and quick start}
\label{sec:installation}

The \texttt{DLScanner} module is available in the Python Package Index (PyPI) repository
and, therefore, can be easily installed via the command line with
\begin{lstlisting}[language=Python]
pip install DLScanner
\end{lstlisting}
This will install all the basic ingredients required for the scanner package,
including important dependencies like \texttt{vegas}, \texttt{matplotlib} and \texttt{scikit-learn},
and their corresponding dependencies.
Due to the complications related to working with TensorFlow and CUDA,
the user is tasked with installing the appropriate version of TensorFlow~\cite{tensorflow_install}
according to the available CUDA library they may or may not have in their working system~\cite{tensorflow_cuda}.

Additionally,
this package is being developed as open source
with the full source code available via \href{https://github.com/raalraan/DLScanner}{GitHub repository}~\cite{githubrepo}.
Testing and trying code fixes before they are submitted to PyPI,
can be done by cloning the repository and installing the package (preferably in a virtual environment)
\begin{lstlisting}[language=bash]
git clone https://github.com/raalraan/DLScanner.git
cd DLScanner
# Recommended: create and activate a virtual environment before the line below
pip install -e .
\end{lstlisting}
this will make the DLScanner module available from anywhere.
We recommend that a virtual environment is created before installing unreleased code with pip.
A simple way of doing this after \texttt{cd DLScanner} is by running
\begin{lstlisting}[language=bash]
# Create a virtual environment
python -m venv .dlscanner-dev
# Activate virtual environment
source .dlscanner-dev/bin/activate
\end{lstlisting}
where the last command activates the virtual environment.
After activating the virtual environment in a shell,
calls to \texttt{pip install} from that shell will only install
packages to that virtual environment without affecting other shells
or environments.
Remember to make sure that the virtual environment is active
before running python code that uses packages from the virtual environment.
There are several ways to create virtual environments,
replace these instructions with the ones corresponding to your workflow.



\subsection{Package structure}

We have structured our code
to keep most units isolated
according to their function,
with other parts of the package
taking care of uniting the components
into the processes we use for scanning.
The structure of our package
and a rough description of their function are given below:

\begin{itemize}
    \item[\textbf{+}] \texttt{DLScanner}: Main entry point.
    \begin{itemize}
        \item \texttt{gen\_scanner}: Generic scanner classes.
        \item[+] \texttt{hep}: Entry point for HEP tools.
        \begin{itemize}
            \item[+] \texttt{MicrOMEGAs}: Entry point for sampling with \texttt{MicrOMEGAs}.
            \begin{itemize}
                \item[-] \texttt{micromegas\_ML}: ML tools related to \texttt{MicrOMEGAs}.
                \item[-] \texttt{utils\_megas}: Utilities for working with \texttt{MicrOMEGAs}.
                \item[-] \texttt{vegas\_S}: internal vegas submodule for \texttt{MicrOMEGAs}.
            \end{itemize}
            \item[+] \texttt{SPheno}: Entry point for sampling with SPheno.
            \begin{itemize}
                \item[-] \texttt{SPheno}: DL tools related to SPheno.
                \item[-] \texttt{utils}: Utilities for working with SPheno.
                \item[-] \texttt{vegas\_S}: internal vegas submodule for SPheno.
                \end{itemize}
        \end{itemize}
        \item[+] \texttt{samplers}: Entry point for samplers.
        \begin{itemize}
            \item[-] \texttt{ML}: ML methods that can be used for sampling.
        \end{itemize}
        \item[+] \texttt{utilities}: Entry point for utilities.
        \begin{itemize}
            \item[-] \texttt{plot}: Plotting utilities.
            \item[-] \texttt{vegas}: VEGAS mapping utilities.
        \end{itemize}
    \end{itemize}
\end{itemize}

It is important to note here that
some submodules could be included in different types of studies:
\begin{itemize}
    \item \texttt{DLScanner.gen\_scanner}:
        Perform a scan of a user defined function using any DL model, including the ones defined in this package.
    \item \texttt{DLScanner.samplers.ML}:
        Take the DL models defined in this package for convenient inclusion in any type of DL applications.
    \item \texttt{DLScanner.utilities}:
        Take advantage of the plotting capabilities or the VEGAS mapping at any point during
        an analysis using \texttt{DLScanner} or in any other application.
\end{itemize}

\subsection{Analyze the result}
A plotting module has been developed to analyze the collected data points. After the scanning process is complete, the points are saved to an external file. The \texttt{plot} module includes various functions for visualizing the distribution of the saved points, such as histograms, scatter plots, and contour plots. These functions can be accessed as follows
\begin{lstlisting}[language=Python]
from DLScanner.utilities.plot import plot_hist, plot_hist_all, plot_contour, plot_scatter
\end{lstlisting}

Two functions have been implemented for generating histograms: one for plotting a histogram of specific variables and another for plotting histograms of all scanned variables. Both functions take as their first argument the full path to a comma-separated file containing the collected data points. In the case of \texttt{plot\_hist()}, an additional argument is required to specify the column number corresponding to the histogram to be plotted.

The scatter plot function takes three arguments: the full path to the data file, the column number of the first variable, and the column number of the second variable to be plotted. Similarly, the contour plot function requires the same three arguments, formatted identically to those for the scatter plot.
For contour plots, the relationship between the two selected variables is smoothed using Gaussian Kernel Density Estimation (KDE). KDE is a non-parametric technique for estimating the probability density function (PDF) of a random variable. It is widely used in data visualization, statistical inference, and smoothing tasks. The Gaussian KDE method achieves smoothing by placing a Gaussian function at each data point and summing these functions to approximate the underlying distribution.
Key advantages of Gaussian KDE include its flexibility, as it does not assume a specific parametric form for the data, and being normalized, which ensures that the estimated PDF integrates to 1.
\section{Sampling methods}
\label{sec:methods}
This section explores the structure of various adaptive sampling methods that can be integrated into the scanning loop to achieve rapid convergence toward the target region. We discuss the currently implemented methods, including DL regressors and classifiers. For DL classifiers, the package supports MLP and SL networks, while only MLP is utilized for regression tasks. Both approaches are augmented by dynamic presampling using VEGAS mapping. This section provides a detailed explanation of the implementation of these methods within the \texttt{DLScanner} package.  

\subsection{DL regressor}
The DL regressor is used  when the goal is to train a model that predicts numerical outputs based on the calculation of observables. For this purpose, the training process requires input parameter values and their corresponding results from HEP calculations. By providing a large set of parameter values, the model generates a set of predictions for the observables, $\hat{Y}$, which can guide the identification of regions of interest. The $K$ set, representing these regions, can be defined using criteria such as $\chi^2$,   or likelihood values. Alternatively, various selection mechanisms can be designed to leverage the model predictions for observables. The accuracy of these predictions is expected to improve with additional iterations, which can be evaluated using metrics such as the mean squared error (MSE).

Over successive iterations, this approach refines the sampling of parameter space, concentrating around regions of interest that inform the construction of the 
$K$ set. In this process the model becomes increasingly specialized, delivering accurate and efficient predictions within these targeted regions.

In this approach, a likelihood can be used to identify the most probable points to fit the constrains. A combined likelihood function can be constructed for multiple constrains as
\begin{equation}
    \mathcal{L}_{\rm combined} =  \prod_{i=1}^N  \rm exp \ \left(-\frac{(Y_i - Y^\text{exp}_i)^2}{2\sigma_i^2} \right)\,,
    \label{eq:likelihood}
\end{equation}
where $N$ is the number of applied constrains used to define the target region, $Y_i$ and $Y^\text{exp}_i$ are the true result and experimental result, respectively, and $\sigma_i$ defines the standard deviation from the best fit value. The output of the likelihood function is a continuous value ranging from 0 to 1. A threshold on the likelihood output is set to consider the refined batch of points, $K$. The selected batch of points also includes a fraction of random points to ensure coverage of the full parameter space. The amount of random points is fixed by the user and can change from one task to another.  The refined points are accumulated to train the network on a larger dataset. Expectedly, The network accuracy increases with the size of the accumulated points.

Sampling with the regressor introduces several challenges. One key issue is that the training sample at each step may lack points sufficiently close to regions of interest. Another concern is that successive training sets can become biased toward specific regions, e.g., as when most of the outputs are similar and far from the desired regions.  
To address the first issue, known maxima can be used as seeds, with additional samples drawn in their vicinity. For the second issue, it is crucial to construct training sets that capture the diversity of possible outputs.
This is where the model predictions can play a pivotal role. By leveraging these predictions, we can identify points that provide a diverse and relevant range of results. These points can then be prioritized for HEP calculations and included in subsequent training steps.  
Over multiple iterations, this approach should result in a parameter space sampling that is increasingly concentrated in the regions of interest, as defined by the $K$ set. Consequently, the model becomes specialized in delivering accurate and efficient predictions for these targeted regions.

The scan results and convergence depends on the structure of the used DL regressor. Focusing on MLP network, consider $(H-1)$ hidden layers and one output layer, with hidden layers structures as 
\begin{equation}
    z_n^{h} = \mathcal{F}\left(W^h z_n^{h-1}+b_n^h \right)\,, \hspace{4mm} h= 1,2,\dots, H-1\,,
\end{equation}
with $z_n^h$ is the output of neuron $n$ in  the $h$ hidden layer, $W^h$ is a weight matrix to be updated during the network training and it has the dimension of $(n_h\times n_{h-1})$, $b$ is a bias vector and $\mathcal{F}(.)$ is a nonlinear activation function, usually ReLU: $\mathcal{F}(x) = \text{ max} (0,x)$. The structure of the output layer is similar to the hidden layer but without the activation function as 
\begin{equation}
    \hat{Y_i} = W^H z^{H-1}_i + b^H_i\,,
\end{equation}
with $\hat{Y}$ is the predicted values of the input. A loss function is computed to quantify the error between the predicted and true labels. The MSE loss is defined as
\begin{equation}
    f_\text{MSE} = \frac{1}{M}\sum_{i=1}^M\left(\hat{Y}_i-Y_i \right)^2\,,
\end{equation}
with $M$ the size of the training batch. To minimize the loss function the weight matrices, $W^{h}$, are updated with new values during the network training process. This can be achieved using back-propagation with optimizers, e.g. the poular Adam optimizer~\cite{kingma2014adam}.
During the training process, the weight matrices are updated in every iteration or ``epoch'', fitting the MLP regressor to 
effectively learn complex relationships between the input and output training dataset.
The trained MLP network is then used to predict new points, $\hat{Y}_i$, to be evaluated with the likelihood function, Eq.~\eqref{eq:likelihood}. 
Network training and prediction process is the step number 2 as mentioned in the Sec.~\ref{sec:terminology}.  This has to be done in each scanning loop with increasing size of the accumulated points, which in turn increases the model ability to model the complex relationship between the input sampled points and the target region.

\subsection{DL classifier}
The objective of the DLScanner approach is to learn a relationship between the sampled points and target boundaries. In this method, binary labels, 0 or 1, are used to indicate whether a point lies outside or inside the target region, respectively. Unlike the regression-based approach, the DL classifier does not rely on likelihood. Instead, it uses a labeling function to evaluate whether sampled points fall inside or outside the target region and assigns binary labels accordingly.

\texttt{DLScanner} comes with two DL classifiers, MLP and SL networks. Both are incorporated with either VEGAS map or random sampling. The structure of the scanning loop in both cases is the same, with the only difference being the structure of the used network.
\subsubsection{Multi-Layer Perceptron}

\begin{table}[tbh!]
\centering
\begin{tabularx}{\textwidth}{@{}X X X@{}}
\toprule
         & \textbf{MLP Classifier}                                           & \textbf{MLP Regressor}                                           \\ \midrule
\text{Task}              & Classification of data into discrete classes.                    & Predicting continuous output values.                            \\ \midrule
\text{Points identification}              & Likelihood free.                    & Likelihood based.                            \\ \midrule

\text{Output}            & A probability distribution over $K$ classes: $\hat{Y}_i \in [0, 1]^K$, where $\sum_k \hat{Y}_{ik} = 1$. & A continuous scalar or vector: $\hat{Y}_i \in \mathbb{R}$.    \\ \midrule
\text{Refined points (K)}              & Chosen based on threshold of the network output.                    & Chosen based on threshold of the likelihood function.                            \\ \midrule
\text{Labels}              & Binary labels: 1 for valid points, 0 for invalid points.                    & Continuous labels with exact value of observables.                            \\ \midrule
\text{Output Layer}      & Uses a sigmoid activation to convert output into probabilities.   & Uses a linear activation (no activation function).              \\ \midrule
\text{Loss Function}     & Cross-entropy loss: $\mathcal{L} = -\frac{1}{M} \sum_{i=1}^M  Y_{i} \log \hat{Y}_{i}$. & MSE: $\mathcal{L} = \frac{1}{M} \sum_{i=1}^M (\hat{Y}_i - Y_i)^2$. \\ \midrule
\text{Objective}         & Maximize the probability of the target class.                   & Minimize the error between predicted and true target values.  \\ \midrule
\text{Boundary or Mapping} & Learns decision boundaries to separate between valid and invalid points.        & Learns a continuous mapping between input and output.  \\ \midrule
\text{Interpretation of Output} & The class with the highest probability is the prediction: $\hat{Y}_i = \arg\max_k \hat{Y}_{ik}$. & The output is the predicted value for the input point: $\hat{y}_i$. \\ \midrule
\text{Decision Boundaries} & Non-linear surfaces partitioning the input space into regions for different classes. & No explicit boundaries. Maps the entire input space to output values. \\ \midrule
\text{Training Focus}     & Adjust weights to correctly classify inputs by adjusting class probabilities. & Adjust weights to fit the predicted values to continuous targets. \\ \midrule
\text{Hidden Layers Role} & Transform features to create separable regions for classes.      & Transform features to capture relationships between input and output. \\ \bottomrule
\end{tabularx}
\caption{Key differences between the role of MLP Classifier and MLP Regressor into the scanning loop. In this table, classes indicate the regions span the valid and invalid points. Input and output points indicate the sampled and predicted points.}
\label{tab:mlp_comparison}
\end{table}


Unlike the regressor described earlier, this approach relies on a HEP calculation that produces a binary output. Structure and training of the MLP classifier is the same as the MLP regressor case with little differences. The structure of the MLP regressor and classifier are the same apart from the output layer, which has one neuron with sigmoid activation function. This function transform the network predictions into probability, $\hat{Y}$, ranging from 0 to 1. Points with $\hat{Y}$ close to 1 are considered as most likely valid points while those with $\hat{Y}$ close to 0 are most likely invalid points. Moreover, minimization of the error between the MLP classifier predictions and true labels is done using Binary Cross Entropy (BCE) with a functional form
\begin{equation}
    f_\text{BCE} = -\frac{1}{M}\sum_i^M\left(Y_i\log(\hat{Y}_i)+(1-Y_i)\log(1-\hat{Y}_i) \right) \,, 
\end{equation}
with $M$ indicating the size of the training batch and $Y_i,\hat{Y}_i$ are the true binary labels and the predicted ones. Similar to the MLP regressor, Adam optimizer can be used to minimize the BCE loss function. Iterating through the training process the MLP classifier learns the properties of both classes, valid and invalid points. 

For this model, predictions are distributed in the range $[0, 1]$, representing a confidence score.
This confidence score enables various strategies for selecting points. A straightforward method is to construct the $K$ set from points most likely to be inside the region of interest, i.e., those with $\hat{Y}$ close to 1. As the goal is to learn the decision boundaries between the valid and invalid points, we accumulate equal size of points with $\hat{Y}\sim 1 $ and $\hat{Y}\sim 0$. These points are considered as the most probable points to represents the valid and invalid points, respectively. In this case the classifier network uses equal size datasets for training. This has two advantages: first it overcomes the problem of imbalanced dataset in which the network overestimate one dataset over the other. Second, the invalid points are not discarded but used to train the network to learn the properties of the invalid points which results accurate predictions the boundaries.


While the goal of the regressor is to specialize in making precise predictions within highly relevant regions, the objective of the classifier is to accurately define the boundary separating the two classes. As a result, the sampled points are expected to concentrate around this decision boundary, reflecting areas where the model has undergone extensive training. Moreover, because training occurs with balanced samples, there may also be a concentration of points in the smaller class, which often corresponds to the region of interest.  Key differences between the role of MLP classifier and MLP regressor in the scanning loop are summarized in table~\ref{tab:mlp_comparison}.
\subsubsection{Similarity learning}
\begin{figure}[h!]
    \centering
    \includegraphics[width=0.9\linewidth]{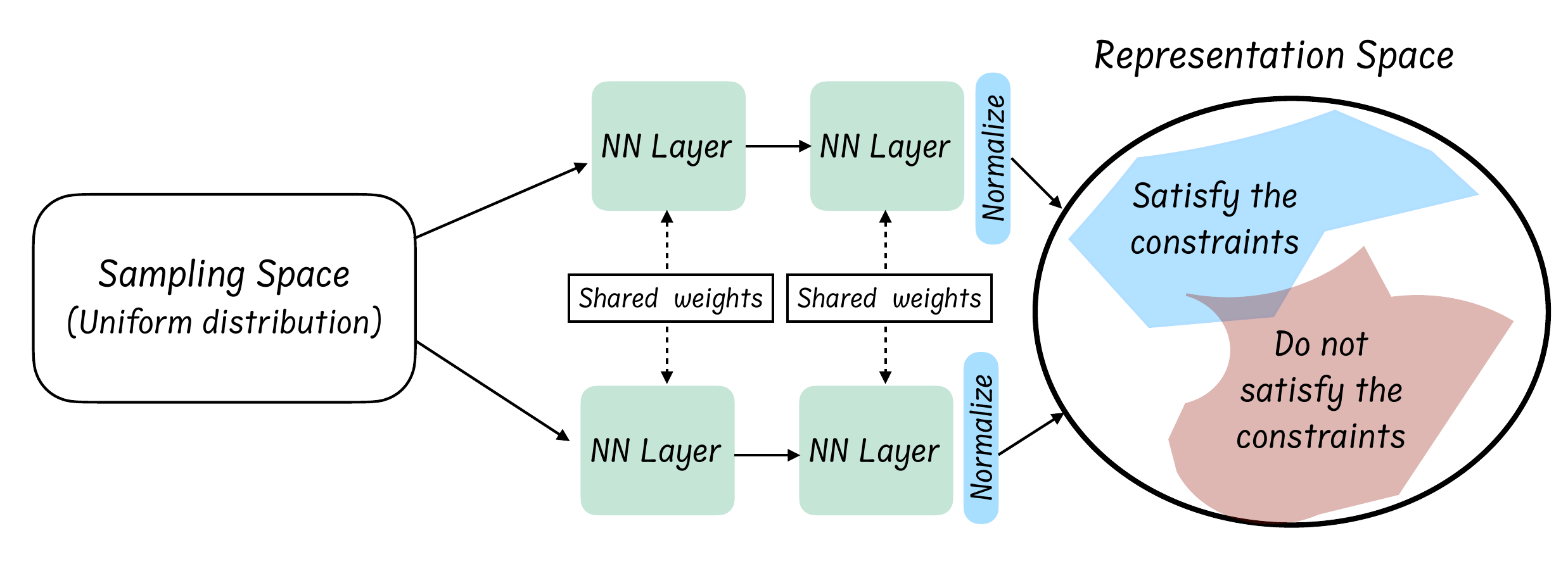}
    \caption{Schematic representation of similarity learning within the scanning loop. During training, the network maps sampled points that satisfy the constraints to a different region of the representation space than those that do not. Once the representation space is structured, it can then be used to predict points that are very close to the target region. }
    \label{fig:SL}
\end{figure}

Supervised SL is a technique used to learn a representation space where similar inputs are grouped close together while dissimilar inputs are moved far apart in a latent space.
The network consists of two training steps: The first step  comprises a network of two identical encoders to map high dimensional input to a low dimensional space, as shown in Fig.~\ref{fig:SL}.
The shared weights are essential because they ensure that the same transformation is applied to both inputs.
The encoders map the input data into a common embedding space, and the similarity between the embeddings is measured using a distance metric like Euclidean distance or cosine similarity.
The loss function then penalizes the network for assigning high similarity to dissimilar inputs and encourages it to bring similar inputs closer in the embedding space.
In the second step, a fully connected layer is added to analyze the mapped data onto the representation space with an additional output layer with two neurons to classify the input.


Given two inputs, $x_1$ and $x_2$, the SL network encodes them using a shared encoder function $f(.)$. The encoder function $f$
could be any deep neural network that outputs a fixed-dimensional vector. After encoding the inputs, the resulting embedding are denoted as:
\begin{equation}
    z_1 = f(x_1), \hspace{6mm} z_2 = f(x_2)\,.
\end{equation}
The similarity between the two embeddings $z_1$ and $z_2$ can be measured by computing their Euclidean distance as 
\begin{equation}
    d_\text{Euc}(z_1,z_2) = \sqrt{\sum_i^n \left (z_1^i - z_2^i \right)^2}\,,
\end{equation}
with $n$ is the dimension of the latent space. In this case similar pairs are grouped close together in the latent space. Alternatively, the similarity can be measures using a  cosine similarity distance as 
\begin{equation}
    d_\text{cs}(z_1,z_2) = \frac{z_1\cdot z_2}{\lVert z_1\rVert\lVert z_2\rVert} = \frac{\sum_i^n z_1z_2}{\sqrt{\sum_i^n \left(z_1^i \right)^2}\sqrt{\sum_i^n \left(z_2^i \right)^2}}\,,
    \label{eq:contrastive}
\end{equation}
with cosine similarity ranging between 0 and 1 for similar and dissimilar pairs, respectively. For supervised task, the network is trained to minimize a contrastive loss  function of the form 
\begin{equation}
    \mathcal{L}_{\rm cont} = \left(1-y \right) d(z_1,z_2)^2 + y \text{ max}\left(0, m- d(z_1,z_2)^2\right)\,, 
\end{equation}
with $d(z_1,z_2)$ any of the metric functions above and $m$ a margin value to control the similarity learning. We set a default value of $m=1$ in the code.
Once the first training step is done, we freeze the weights of one of the encoders and a fully connected layer is added to reduce the mapped data onto the representation space. An additional output layer is added with two output neurons for the classification. For this training step the weights of the fully connected layer are updated using a cross entropy loss function. 

Supervised SL, with weight-sharing encoders, has several advantages over the MLP classifier. One of the primary advantages is that the SL network architecture focuses on learning the relationships between inputs, rather than simply classifying individual inputs. This makes it well-suited for tasks like verification, where determining whether two inputs are the same or different is crucial. In contrast, MLPs are typically designed for classification tasks, where the goal is to assign an input to a specific class.
Another advantage of supervised similarity learning is that it is more data-efficient. In MLP-based classification tasks, a large amount of labeled data is required for each class, and the model struggles to generalize to new, unseen data. Supervised SL, on the other hand, is capable of generalizing better to unseen data because it learns to compare inputs based on their features, rather than memorizing class-specific patterns.

The structure of the scanning loop is the same as the MLP classifier, only replacing the MLP network with the SL one.
To access the similarity learning network, the function \texttt{similarity\_classifier()} is defined inside the \texttt{DLScanner.samplers.ML} module in which the inputs are the sampled points together with the labels with $Y=1$ and $Y=0$ indicating points inside and outside the target, respectively. The function returns a trained similarity network that can be used to predict new points. All other input arguments are set to their default values and can be changed by the user depends requirements specific to performed study. The default values can be printed out using the following call
\begin{lstlisting}[language=Python]
print(similariy_classifier.__defaults__)
\end{lstlisting}

\section{Scanning tools}
\label{sec:scanning}
In this section, we discuss the integration of the proposed sampling methods within the scanning loop. In the current version, we provide three examples: scanning over any given numerical function and scanning the parameter space of theoretical models using the SPheno and MicrOMEGAs packages. In addition, we show how to incorporate custom numerical packages into the scanning loop which increases the generality of the proposed package beyond the HEP analysis. 
\subsection{Generic scan}

\label{sec:genericscan}

We begin by describing the interface to perform an scan for a user-defined function
over a parameter space.
Later, we will describe how this can be used
to implement scanning on additional tools
not considered here.

The interface for the generic scan is contained in \texttt{DLScanner.gen\_scanner}.
The main class to perform the actual scan is in the class \texttt{sampler}.

\begin{itemize}
    \item \texttt{DLScanner.gen\_scanner.\textbf{sampler}}
    \begin{itemize}
        \item
            \texttt{user\_fun}: callable, function that will be sampled.
        \item
            \texttt{ndim}: integer, number of dimensions that will be scanned over.
        \item
            \texttt{limits}: array-like, boundaries of the parameters being sampled.
        \item
            \texttt{outdim=1}: integer, optional, number of outputs given by \texttt{user\_fun}.
        \item
            \texttt{method="Classifier"}: string, optional,
            choose between using classifier (\texttt{"Classifier"}),
            regressor (\texttt{"Regressor"}) or a network provided by the user (\texttt{"Custom"}).
        \item
            \texttt{model=None}: \texttt{keras.model} object, optional, if the user is providing their own model.
            At the moment only Keras models are used (through TensorFlow).
        \item
            \texttt{optimizer="Adam"}: string or \texttt{keras.optimizers}, optional, optimizer using during training of the network.
        \item
            \texttt{loss=None}: string or callable, optional, loss function that will be used by the optimizer.
            For a user defined function, it must be a function taking two arguments: true and predicted labels.
        \item
            \texttt{sample0=None}: array-like, optional, an initial sample to be used for training.
            This is for cases where some data is already present.
            If not given, it will be created before the initial training.
        \item
            \texttt{out0=None}: array-like, optional, output corresponding to \texttt{sample0}.
            This is for cases where some data is already present.
            If not given, it will be created before the initial training.
        \item
            \texttt{K=100}: integer, optional,
            number of points that will be selected at every step from network prediction
            to be passed to \texttt{user\_fun}.
        \item
            \texttt{randpts=None}: integer, optional,
            number of random points that will be added to points suggested according to network prediction.
            Mostly for discovery of new regions.
            If not given, it will be a fraction of 0.1 of \texttt{K}.
        \item
            \texttt{L=None}: integer, optional,
            number of random points generated from trained VEGAS map and/or random distribution,
            that will be passed directly to network for prediction.
            If not given, it will be 100 times \texttt{K}.
        \item
            \texttt{neurons=100}: integer, optional,
            number of neurons of hidden layers in network.
        \item
            \texttt{hlayers=4}: integer, optional,
            number of hidden layers in network.
        \item
            \texttt{learning\_rate=0.001}: float, optional,
            learning rate to be used for the optimizer.
        \item
            \texttt{epochs=1000}: integer, optional,
            number of epochs that will be used for each training.
        \item
            \texttt{batch\_size=32}: integer, optional,
            size of batches used during training.
        \item
            \texttt{verbose=1}: integer, optional,
            Amount of information that will be shown during running.
            0 means no output, 1 shows minimal information about steps,
            2 shows information about training.
        \item
            \texttt{use\_vegas\_map=True}: boolean, optional,
            Whether or not to train and use a VEGAS map to generate test points.
        \item
            \texttt{vegas\_frac=None}: float, optional,
            Fraction of \text{L} that will be generated using the vegas map.
            The rest of \text{L} will be generated from random distribution.
        \item
            \texttt{seed=42}: integer, optional,
            seed for the random number generator used internally.
        \item
            \texttt{callbacks=None}: list, optional,
            list of callbacks used during training of the network.
            Only keras callbacks are supported at the moment.
    \end{itemize}
\end{itemize}

This class can easily be used to start a scan on a user defined function.
A very simple example is given below.

\begin{lstlisting}[language=Python]
from DLScanner.gen_scanner import sampler
from user_module import user_function

ndim = 3
limits = [[-10, 10]]*ndim
method = "Classifier"
vegas_frac = 0.5

my_sampler = sampler(
    user_function, ndim, limits, method, vegas_frac=vegas_frac
)
\end{lstlisting}

This starts the sampler instance and also takes care of the 0$^\text{th}$ training
with the first set of random points.
To actually train the network iteratively and collect some points
one also needs to \emph{advance} the process with
\begin{lstlisting}[language=Python]
steps = 10
my_sampler.advance(steps)
\end{lstlisting}
this will run the iterative process for 10 steps.
This iterative process is described in Fig.~\ref{fig:chart}.
After the 10 steps are finished,
the user can easily check the points that have been collected in
\texttt{sampler.sample}
and their corresponding output for \texttt{user\_function}
in \texttt{sampler.sample\_out}.
Analysis can be carried on the obtained sample and its output,
and, if more samples are required,
it is enough to call \texttt{sampler.advance(more\_steps)}
for a previously defined (\texttt{more\_steps}) number of extra steps.
\begin{lstlisting}[language=Python]
more_steps = 20
my_sampler.advance(more_steps)
\end{lstlisting}
This will start where the previous \texttt{advance()}
call ended and keep collecting points while also training the network.
The call to \texttt{advance()} can receive more arguments described below.
\begin{itemize}
    \item \texttt{DLScanner.gen\_scanner.sampler.\textbf{advance}}
    \begin{itemize}
        \item
            \texttt{steps=1}: integer, optional,
            number of steps to run the iterative process.
        \item
            \texttt{epochs=None}: integer, optional,
            number of epochs that will be used for each training.
            If not given, it is taken from sampler instance.
        \item
            \texttt{batch\_size=None}: integer, optional,
            size of batches used during training.
            If not given, it is taken from sampler instance.
        \item
            \texttt{verbose=None}: integer, optional,
            Amount of information that will be shown during running.
            0 means no output, 1 shows minimal information about steps,
            2 shows information about training.
            If not given, it is taken from sampler instance.
        \item
            \texttt{callbacks=None}: list, optional,
            list of callbacks used during training of the network.
            Only keras callbacks are supported at the moment.
            If not given, it is taken from sampler instance.
    \end{itemize}
\end{itemize}

A summary of useful attributes of the \texttt{sampler} is given below:
\begin{itemize}
\item \texttt{DLScanner.gen\_scanner.{sampler}} useful attributes
    \begin{itemize}
        \item
            \texttt{.advance}:
            Run the iterative process for agiven number of steps.
        \item
            \texttt{.sample}:
            Collected samples while running the iterative process.
        \item
            \texttt{.sample\_list}:
            Same as above, but separated by step.
        \item
            \texttt{.sample\_out}:
            Output for collected samples for the function given in \texttt{user\_fun}.
        \item
            \texttt{.sample\_out\_list}:
            Same as above, but separated by step.
        \item
            \texttt{.histories}:
            Histories for the trainings performed in each step.
        \item
            \texttt{.model}:
            Trained model used to suggest new points.
        \item
            \texttt{.vegas\_map\_gen}:
            Random sample generator that uses the last trained VEGAS map 
            to generate a given number of samples.
            Use it as \texttt{.vegas\_map\_gen(n)}.
            Returns \texttt{n} samples and the corresponding Jacobian.
    \end{itemize}
\end{itemize}

In Appendix~\ref{sec:genexample} we show an example that showcases the usage of the \texttt{sampler} class and its attributes.
To illustrate in this section the way the process works and the results obtained
we perform a scan on a 3-dimensional space for the following function
\begin{equation}
    \label{eq:3dmodel}
    O_\text{3d}(x_1, x_2, x_3) = \left[2 + \cos\left(\frac{x_1}{7}\right) \cos\left(\frac{x_2}{7}\right) \cos\left(\frac{x_3}{7}\right)\right]^5\,.
\end{equation}
The true classifier is defined by
\begin{equation}
    \label{eq:3dclass}
        f_\text{class} =
    \begin{cases}
        0 & \text{if}\quad \exp \left(-\frac{1}{2}\left( \frac{O_\text{3d} - 150}{5} \right)^2
        \right) \leq 0.5 \\
        1 & \text{if}\quad \exp \left(-\frac{1}{2}\left( \frac{O_\text{3d} - 150}{5} \right)^2
        \right) > 0.5 \\
    \end{cases}
\end{equation}
which is used as \texttt{user\_function} in the code in Appendix~\ref{sec:genexample}.

\begin{figure}[tb!]
   \centering
    \includegraphics[width=0.5\textwidth]{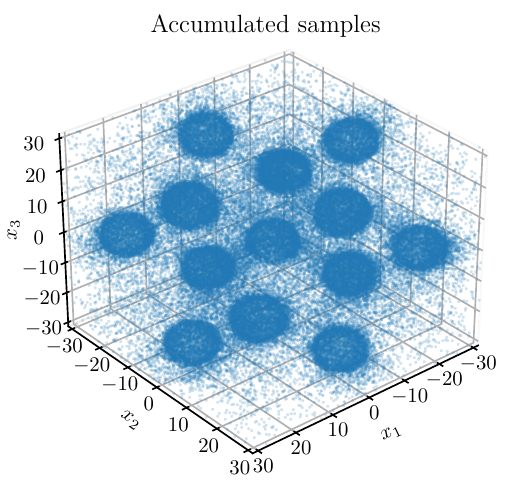}%
    \includegraphics[width=0.5\textwidth]{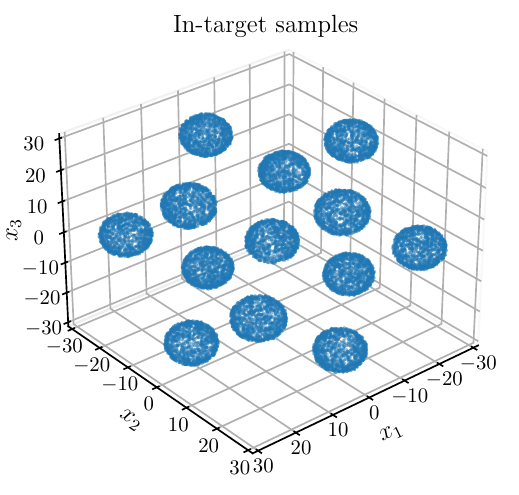}
   \caption{%
       Collected samples for the function $O_\text{3d}$ as described in the text.
       In the left, the total of samples collected in and out of target are displayed (90\,000 points, opacity of each point is 0.2).
       In the right, only the samples collected in target are displayed (17\,000 points, opacity of each point is 0.5).
       The points in the left correspond to the points that were selected by the network and ultimately passed to $O_\text{3d}$.
   }
    \label{fig:inoutsamples}
\end{figure}

\begin{figure}[b!]
   \centering
    \includegraphics[width=0.33\textwidth]{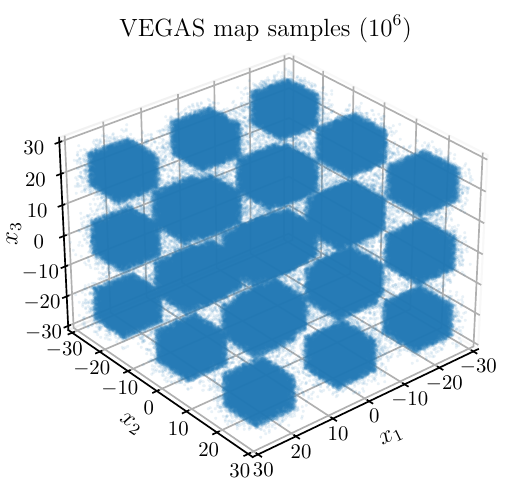}%
    \includegraphics[width=0.33\textwidth]{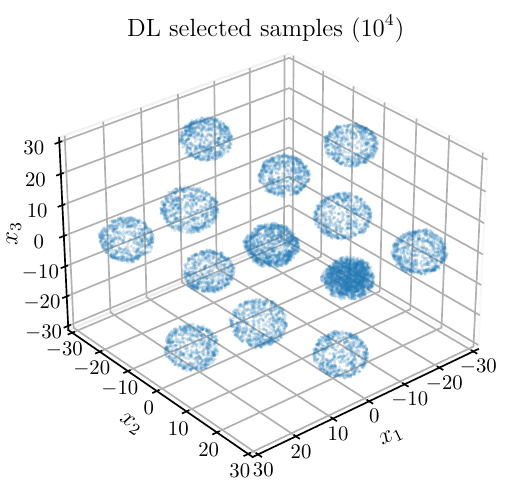}%
    \includegraphics[width=0.33\textwidth]{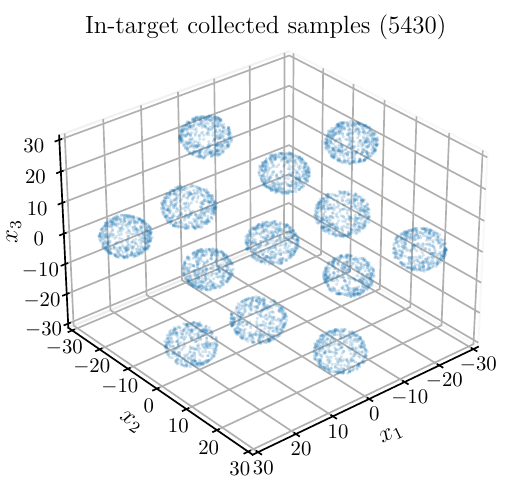}
   \caption{%
       Progression of reduction of samples in the process used here.
       The leftmost panel corresponds to the distribution of points generated when using VEGAS map
       for a total of $10^6$ points (opacity of each point is 0.1).
       The panel in the middle corresponds the filter using the network
       to predict which points are either in- or out-target
       and selecting only $10^4$ points (opacity of each point is 0.2).
       The rightmost panel corresponds to the final points that would be collected as in-target
       during the step, in this case 5430 points (opacity of each point is 0.2).
   }
    \label{fig:sampleprog}
\end{figure}

In Fig.~\ref{fig:inoutsamples} we show a scatter plot of the points accumulated for 8 advance steps
with $K=10^4$ and $L=10^6$.
In the left panel we can see that the points are mostly concentrated in the target regions
that can be seen in the right panel where we show only points in target.
The points shown in the right correspond to the points that were passed to
$O_\text{3d}$ for calculation of true values.
To illustrate the use of VEGAS,
in Fig.~\ref{fig:sampleprog} we show the progression of sample creation and selection
with the points generated by the VEGAS map in the left panel,
the points selected by the network in the middle panel
and the points ultimately checked to be in-target after passing the points in the middle to $O_\text{3d}$.

In Sec.~\ref{sec:newtool} we comment on how this generic scanning class
can be used to implement scanning on new tools
with some assistance by the user.

\subsection{SPheno}
SPheno~\cite{Porod:2011nf,Porod:2003um} is a software tool designed to precisely compute the supersymmetric particle spectrum within high energy scale frameworks, including minimal supergravity, gauge-mediated supersymmetry breaking, anomaly-mediated supersymmetry breaking, and string-effective field theories. It features a user-friendly interface to facilitate the incorporation of other high energy scale models. The program numerically solves renormalization group equations up to two-loop order, allowing users to define custom boundary conditions. 

In order to use the SPheno scan, the user has to install the SPheno package with the required theoretical model. The binary executable file   should be located in the \texttt{bin} directory as \texttt{SPhenoMSSM}, for the MSSM model. An input file in the LesHouches format is used to adjust the numerical values for a single point. This file has to placed in the main directory of the SPheno package. 

\texttt{DLScanner} comes with three scanning modules for different DL methods, when using the SPheno package. The function \texttt{MLP\_regressor()} which utilizes MLP regressor network into the scanning loop. The other functions \texttt{MLPC()} and \texttt{ML\_SL()} utilizes MLP and SL classifier networks, respectively. These functions are implemented in the \texttt{DLScanner.hep.SPheno} module  with the following default arguments:
\begin{itemize}
    \item \texttt{vegas=True}: Boolean to decide between use VEGAS (\texttt{True}) or random sampling (\texttt{False}) to suggest points for the trained DL for predictions. 
    \item \texttt{collected\_points=5000}: Number of the required collected points in the target region.
    \item \texttt{L1=100}: Initial collected in the target region from a random scan to train the network. This is the $0^\text{th}$ step and is used only once to start the scanning loop.
    \item\texttt{L=1000}: Number of the randomly generated points either using VEGAS map (if \texttt{vegas=True}) or from random distribution to be passed to the DL network for prediction.
    \item\texttt{K=300}: Number of points to be refined by Spheno.
    \item\texttt{frac=0.2}: Fraction of the random points added when training the DL network at each iteration.
    \item\texttt{learning\_rate=0.01}: Initial value of the learning rate during the network training.
    \item\texttt{num\_FC\_layers=5}: Number of the fully connected layers to be for the used DL network. 
    \item\texttt{neurons=100}: Number of neurons in each of the fully connected layers.
    \item\texttt{print\_output=True}: Boolean to print out the status of the scan during each iteration. 
\end{itemize}
In both, DL classifiers and regressor, the output layer is fixed to one neuron with \texttt{sigmoid} activation function and \texttt{linear} activation function, respectively. The output of the DL classifiers is in the range $\hat{Y}_{\rm class}=[0,1]$ indicating that the sampled point is on or off-target. The output of the DL regressor is a continuous value of a likelihood function $\hat{Y}_{\rm reg}=[0,1]$, with $\hat{Y}_{\rm reg}\sim 1$ being most likely closer to the best fit value.

For all the scanning methods, the user can use one of the scanning functions as 
\begin{lstlisting}[language=Python]
from DLScanner.hep.SPheno.SPheno import ML_SL
ML_SL()
\end{lstlisting}
when executing the scanning function, the user receives a message asking to provide the path to an input file:
\begin{lstlisting}[language=bash]
>> Please enter the full path to the input file (including the file name): 
\end{lstlisting}
The input file has a specific format in which the user defines the ranges of the scanning parameters as well as the target conditions as follows.

\begin{lstlisting}[language=bash]
pathS: /home/SPheno-4.0.3
Lesh:  LesHouches.in.MSSM_low_2
SPHENOMODEL: MSSM    
output_dir: /home/output_MSSM   
###################################
### Define the scan values        #
###################################
TotVarScanned: 3

VarMin: 1.00000E+02  
VarMax: 1.5000E+04   
VarLabel: # M1input     
VarNum:  1

VarMin: 1.00000E+02
VarMax: 1.5000E+04    
VarLabel: # M2input  
VarNum:  2

VarMin: -1.500000E+03   
VarMax: 1.500000E+03
VarLabel: # Muinput        
VarNum:  23

###################################
### Define the traget region      #
###################################
TotTarget: 2

TargetMin: 124  
TargetMax: 126  
TargetLabel: # hh_1 
TargetNum: 25 
TargetResNum: 1

TargetMin: 300  
TargetMax: 900  
TargetLabel: # hh_2 
TargetNum: 35 
TargetResNum: 1

\end{lstlisting}

The extension of the input file is not important, e.g., input.txt, input.dat, etc, work all the same.
The following information has to be provided inside the input file:
\begin{itemize}
    \item \texttt{pathS}: Full path to the SPheno directory.
    \item\texttt{Lesh}: Name of the LesHouches file, Spheno input. This file has to be in the main SPheno folder.
    \item \texttt{SPHENOMODEL}: Name of the used executable model in the SPheno bin directory.
    \item \texttt{output\_dir}: Name of the directory to store the output points. If it does not exist it will be created. 
    \item \texttt{TotVarScanned}: Number of the scanned variables. 
    \item \texttt{VarMin}: Minimum value of the scanned variable.
    \item \texttt{VarMax}: Maximum value of the scanned variable.
    \item \texttt{VarLabel}: Name of the variable as written in the LesHouches file.
    \item \texttt{VarNum}: LesHouches  number of the scanned variable as written in the input LesHouches file.
    \item \texttt{TotTarget}: Number of constraints to be satisfied.
\end{itemize}

The labels in the Target block must match exactly as they appear in the SPheno output, as the package processes the output by reading it line by line to extract the constraints. Once the required number of in-target points is reached, the values of the scanned points are saved in a file named \texttt{Accumulated\_points.txt}, located in the output directory specified in the input file. Additionally, the output spectra from SPheno for the in-target points are also stored in the output directory. These spectra provide valuable information, as they include additional details about the theoretical model. 

For the DL regressor, the user must specify a likelihood threshold. The network will then accumulate points that exceed this threshold. Additionally, \texttt{DLScanner} includes a trial SPheno code to reproduce the results presented in Section 6, achieved using the SL classifier with VEGAS. This can be accomplished using the following function:
\begin{lstlisting}[language=Python]
from DLScanner.hep.SPheno.SPheno_trial import ML_SL_trial
ML_SL_trial()
\end{lstlisting}
Once executed the code will automatically download locally the binary file for the MSSM\footnote{This requires GLIBCXX\_3.4.29.}  and the input file to reproduce the results.
\subsection{micrOMEGAs}
The \texttt{micrOMEGAs}~\cite{Alguero:2023zol} package is a computational tool designed to analyze the properties of cold dark matter within a general particle physics framework. Initially developed to determine the relic density of dark matter, it has been extended to calculate predictions for both direct and indirect detection of dark matter. For a single point computation, \texttt{micrOMEGAs} requires an input parameter file with the numerical values of the computed point. To scan over the free parameters in the input file using MLP classifier, the following code can be used
\begin{lstlisting}[language=Python]
from DLScanner.hep.MicrOMEGAs import micromegas_ML
micromegas_ML()
\end{lstlisting}

The function \texttt{micromegas\_ML()} has the same default parameters as the MLP classifier function for \texttt{SPheno} scan, discussed earlier. Once the above code is executed it asks for an input file. The input file for  \texttt{micrOMEGAs} scan has the following structure
\begin{lstlisting}[language=Python]
path_micromegas: /home/micromegas_6.0.4/MSSM
input: mssm1.par
output_dir: /home/output_mOmegas         
###################################
### Define the scan values        #
###################################
TotVarScanned: 2

VarMin: 2.00000E+02  
VarMax: 4.000E+02   
VarLabel: mu     


VarMin: 5.00000E+00
VarMax: 3.000E+01    
VarLabel: tb
###################################
### Define the traget region      #
###################################
TotTarget: 1

TargetMin: 0.09
TargetMax: 0.1  
TargetLabel: Omega
\end{lstlisting}

Similar to the \texttt{SPheno} case, the extension of the input file is irrelevant.
The following information has to be provided inside the input file:
\begin{itemize}
    \item \texttt{path\_micromegas}: Full path to the working directory that contains the executable  \texttt{main} file.
    \item\texttt{input}: Name of the input file for the single point calculation. This file has to be in the same working directory.
    \item \texttt{output\_dir}: Name of the directory to store the output points. If the file does not exist the code will create one. 
    \item \texttt{TotVarScanned}: Number of the scanned variables. 
    \item \texttt{VarMin}: Minimum value of the scanned variable.
    \item \texttt{VarMax}: Maximum value of the scanned variable.
    \item \texttt{VarLabel}: Name of the variable as written in the input file.
    \item \texttt{TotTarget}: Number of constraints to be satisfied.
\end{itemize}

At each iteration the code runs the input file, for the single point, with the new sampled values for the scanned parameters and directs the output to a virtual output file. To check the constraints the code searches for the float after the given text and check if this value is between \texttt{TargetMin} and \texttt{TargetMax}. 
\subsection{Adding new tool}
\label{sec:newtool}

Here we comment briefly on how to perform a scan on additional tools required by the user.
The point is to get the new tool output into a function that can take an array of vector of the parameter space
and output an array of the output for each vector.
Here we will consider the most complicated case:
A tool that must be called from the command line and outputs text that must be parsed.
This is the case for several HEP tools that are written in C/C++ and/or Fortran and
need to be compiled before running the calculation.
In python, two things are required to first get the numerical result as a python float:
execute the command catching the output
and parsing the output down to the required numerical value.
In python, \texttt{Popen} from the \texttt{subprocess} module is used to run executables
and catch their output.
Then, the user must take care of parsing the output down to the desired numerical values
\begin{lstlisting}[language=Python]
import numpy as np
from subprocess import Popen, PIPE
# It is assumed that the user knows how to parse the output of the program
# and, for the classifier, that has decided on a condition for points that
# are in- and out-target
from user_module import parse_my_output, write_parameters, user_condition

my_program = "./my_executable"
# If parameters are read from file
parameters_file = "./my_parameters"

# Function to run program and parse content.  Parameters read from command line arguments
def run_my_program_1(pvector):
    par1, par2, par3 = pvector
    process = Popen([my_program, par1, par2, par3], stdout=PIPE, stderr=PIPE)
    output, error = process.communicate()
    return parse_my_output(output)  # Parsing returns only numerical value

# Function to run program and parse content.  Parameters read from file
def run_my_program_2(pvector):
    write_parameters(parameters_file, pvector)
    process = Popen([my_program, parameters_file], stdout=PIPE, stderr=PIPE)
    output, error = process.communicate()
    return parse_my_output(output)  # Parsing returns only numerical value

# Function to take an array of parameter vectors and output array of output
def run_array(array):
    result = np.empty(len(array))
    for j in range(len(array)):
        result[j] = run_my_program(pvector[j])
    return result  # Output an array of calculation results

# For the classifier: function that separates into classes: in- and out-target
def true_class(array):
    result = run_array(array)
    labels = user_condition(result)
    return labels  # Array of 0 and 1 values
\end{lstlisting}
This module can be imported together with \texttt{DLScanner}
and either \texttt{run\_array} (regressor)
or \texttt{true\_class} can be used as the \texttt{user\_fun}
in Sec.~\ref{sec:genericscan},
also setting the appropriate limits and number of dimensions.
In the example above the number of dimensions is 3 but it should be easy
to generalize to more.
It is here where parallelization must be implemented for the tool,
in this case in the \texttt{run\_array()} function.

\section{MSSM example}
\label{sec:mssm}
In this section we use the presented scanning tools to study the MSSM parameter space using the SPheno package. We focus on DL scanner tools using SL classifier, MLP classifier and MLP regressor. Also, we consider two cases for  sampling using adaptive mapping with VEGAS and random  sampling.

The full scalar potentional of the MSSM is defined as~\cite{Djouadi:2005gj}
\begin{equation}
\begin{split}
    V_H =& \left(|\mu|^2+m^2_{H_1} \right) \left|H_1 \right|^2 + \left(|\mu|^2+m^2_{H_2} \right) \left|H_2 \right|^2 - \left(B_\mu H_1 H_2 + \rm h.c.\right) \\
    & \frac{g^2_1+g^2_2}{8} \left(|H_1|^2-|H_2|^2 \right)^2+\frac{1}{2}g_2^2|H_1^\dagger H_2|^2 \,,
    \end{split}
\end{equation}

with $H_1,H_2$ the two Higgs doublets. The $\mu$ mass parameter for the Higgs-Higgsino and the bilinear tearm for soft SUSY breaking $B_\mu$ are determined by using minimization conditions as
\begin{equation}
    \begin{split}
        \mu^2 &= \frac{m^2_{H_2} \sin^2\beta -m^2_{H_1} \cos^2\beta}{\cos 2\beta}- \frac{m^2_Z}{2}\,,\\
        B_\mu &= \frac{(m^2_{H_1}-m^2_{H_2})\tan 2\beta + m^2_Z\sin 2\beta}{2} \,,
    \end{split}
\end{equation}
with $\tan\beta =v_2/v_1$.
After expansion of the scalar potential in terms of neutral and charged components of the Higgs doublet we find the potential
\begin{equation}
    \begin{split}
        V_H =& m_1^2(|H^0_1|^2+|H^-_1|^2) + m_2^2 (|H^0_2|^2+|H^+_2|^2)+ m_3^2 (H_1^0H_2^0 - H_1^-H_2^+ + \rm h.c.)\\
        &\frac{g_1^2+g_2^2}{2}(|H^0_1|^2+|H^-_1|^2- |H^0_2|^2-|H^+_2|^2)^2 +\frac{g_2^2}{2} |H_1^0H_1^{-*}+H_2^+H_2^{0*}|^2 \,,
    \end{split}
\end{equation}
with $H^0_j,H^\pm_j$ neutral and charged components of the Higgs doublets and the mass terms corresponding to
\begin{equation}
    m_1^2 = |\mu|^2+m^2_{H_1}\,, \hspace{4mm} m_2^2 = |\mu|^2+m^2_{H_2} \,, \hspace{4mm} m_3^2 = B_\mu \,. 
\end{equation}
We consider a scan over $5$ parameters to satisfy the SM-like Higgs measured mass to be in the range $[124-126]$~GeV. The scan ranges are defined as 
\begin{equation}
    \begin{split}
         &100 \ {\rm GeV}\ \le m_1 \le 10^{4}\ \rm GeV \,, \hspace{4mm} 100\ {\rm GeV}\ \le m_2 \le 10^{4}\ \rm GeV\,,  \\ \\
         100 \ \rm GeV\ \le & \ m_3 \le 10^{4}\rm GeV \,, \hspace{4mm}-1500\ \rm GeV\ \le \mu \le 1500\  \rm GeV \,, \hspace{4mm} 5 \le \tan\beta \le 50\,,
    \end{split}
\end{equation}
with masses considered in a diagonal base with $m_l=1000$~GeV and $m_d=m_u=2000$~GeV.

SPheno is used to compute the physical mass of the Higgs boson and we utilize our approach to pinpoint the region with $m_{h_{\rm SM}} = 124\text{-}126$ GeV. For all methods we examine the case of using VEGAS mapping to suggest points to the DL networks and the case with simple random suggestion. 

\subsection{Networks structure}
\label{App:A}
For this scan three  DL networks (SL classifier, MLP classifier and MLP regressor) are used with both VEGAS and random sampling. VEGAS training is set for all the three networks to internally use 5 iterations to refine the mapping with 100 increments and a damping parameter $\alpha=1$. 

For the structure of the MLP classifier and regressor, we employ a MLP architecture consisting of fully connected layers. The network is designed with 5 hidden layers, each containing 100 neurons and utilizing the ReLU activation function. The output layer consists of a single neuron with a linear activation function and a sigmoid activation function for regression and classification tasks, respectively.
As for the loss function, we use the MSE for regression and BCE for classification. In both cases, the model is optimized using the Adam optimizer, configured with a learning rate of 0.001 and exponential decay rates $\beta_1=0.9$ and $\beta_2=0.999$. Training is conducted over 1000 epochs. To evaluate the DL regressor, the $K$-set is constructed by selecting 90\% of the points where the likelihood exceeds 0.9, based on the predictions, $\hat{Y}$,  generated by the DL model. The remaining 10\% of the $K$ set is sampled randomly from the parameter space. During training, we iteratively incorporate the accumulated points into the training set every two iterations, as well as in the final iteration, to enhance the model accuracy.
When using the regressor, all suggested points are accumulated, focusing on the region of interest. In this context, the region of interest corresponds to points where the likelihood exceeds the specified threshold of $0.9$. This iterative process ensures that the model progressively refines its predictions around the targeted region.

For the classifier, the DL outputs a probability representing the model confidence that a given point belongs to the target class. Using this confidence score, $\hat{Y} \in [0, 1] $, we select points from the larger set $L$ to create a smaller subset $K$.  

\begin{table}[!th]
    \setlength\tabcolsep{0.29cm}
    \begin{tabular}{lccc}
        \toprule
         & MLP Regressor & MLP Classifier& SL Classifier \\
        \midrule
        Size of initial set & 100 points & 100 points& 100 points \\
        Size of random test set ($L$) & 50\,000 points &  50\,000 points&  50\,000 points \\
        Size of selection batch($K$) & 300 points & 300 points & 300 points \\
        Random points in $K$ & 10\% & 10\%& 10\% \\
        Input layer (IL) dimension & 5 & 5 & 5 \\
        \textbf{Hidden layers (HL)} & 4 & 4 & 3$\times$2\\
        HL neurons & 100 & 100& 100 \\
        \textbf{Projection layer neurons}& $-$ & $-$ &500\\
        HL activation function & ReLU & ReLU& ReLU \\
        Output layer (OL) dimension & 1 & 1& 1 \\
        \textbf{OL activation function} & linear &sigmoid& linear \\
        \textbf{Loss function} & MSE& BCE&Contrastive loss \\
        Optimizer & Adam  & Adam & Adam \\
        Learning rate & 0.001 & 0.001& 0.001 \\
        Epochs & 500 & 500 & 500 \\
        \bottomrule
    \end{tabular}
    \caption{\label{tab:structure}%
        Hyperparameters of the different DL networks used for the results displayed in Fig.~\ref{fig:performance}.
         }
\end{table}

Similar to the approach with the regressor, $10\%$ of the points in $K$ are selected randomly. For the remaining $90\%$, half  are chosen from points where $\hat{Y} > 0.75$, indicating high confidence that the points are inside the class, and the other half from points where $\hat{Y} \leq 0.75$, including points that are uncertain or likely outside the class. The threshold of $0.75$ is not unique but serves as a convenient choice to split the points into confident and less confident categories.  

As with the regressor, this subset $K$ is passed to the true classification calculation. Points that belong to the class, where $\mathcal{L} > 0.9$ are accumulated for further training and, unlike the regressor, points below the threshold are retained for training purposes.  
It is also crucial to focus the training on two specific categories of points: those that were misclassified by the model and those where $\hat{Y} \approx 0.5$ ,indicating high uncertainty. Including these points in the training process helps the network better refine its decision boundary and improve overall classification accuracy.

For the SL scan, all parameters are fixed to be the same as the MLP classifier. The only difference comes from the different training procedure between the MLP and SL. 
Training of the SL network consists of more steps. First, the network consists of two MLP encoders, that share their weights, with three hidden layers and one projection layer. The three hidden layers consist of $100$ neurons with ReLU activation, while the projection layer has $500$ with linear activation. Output from the two encoders is used to compute the Euclidean distance  and the error is computed by using the contrastive loss function, Eq.~\eqref{eq:contrastive}. The network is trained for 500 epochs and  Adam optimizer is used to minimize the loss function. Once the network is trained, the weights of the first encoder is frozen and a linear fully connected layer with 100 neurons is added with output layer of one neuron and sigmoid activation. The linear layer is trained for 500 epochs and Adam optimizer is used to minimize a BCE loss function. A selected batch, $K$, from SL output is used to be refined using the SPheno package. The loop is then continue similar to the MLP classifier. The structure of the three networks and the values of the used parameters are summarized in table~\ref{tab:structure}.

\begin{figure}[tbh!]
   \centering
    \includegraphics[width=0.9\linewidth]{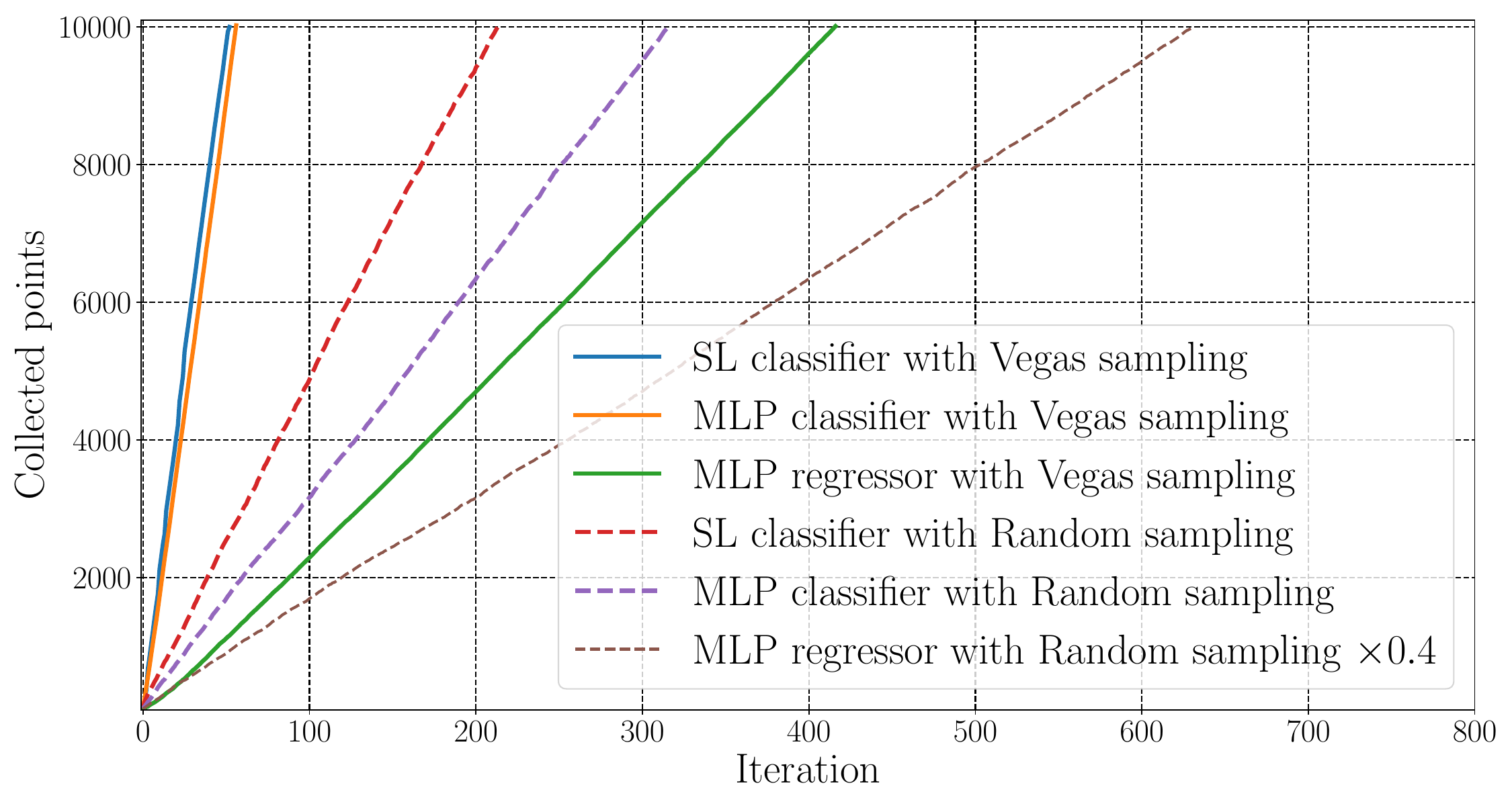}\\
   \caption{Number of accumulated valid points in terms of number of iterations for different scanning methods. Solid lines represent the VEGAS sampled scan while dashed lines represent the random sampling.}
    \label{fig:performance}
\end{figure}

\subsection{Scan result}

\begin{figure}[tbh!]
    \includegraphics[width=\linewidth]{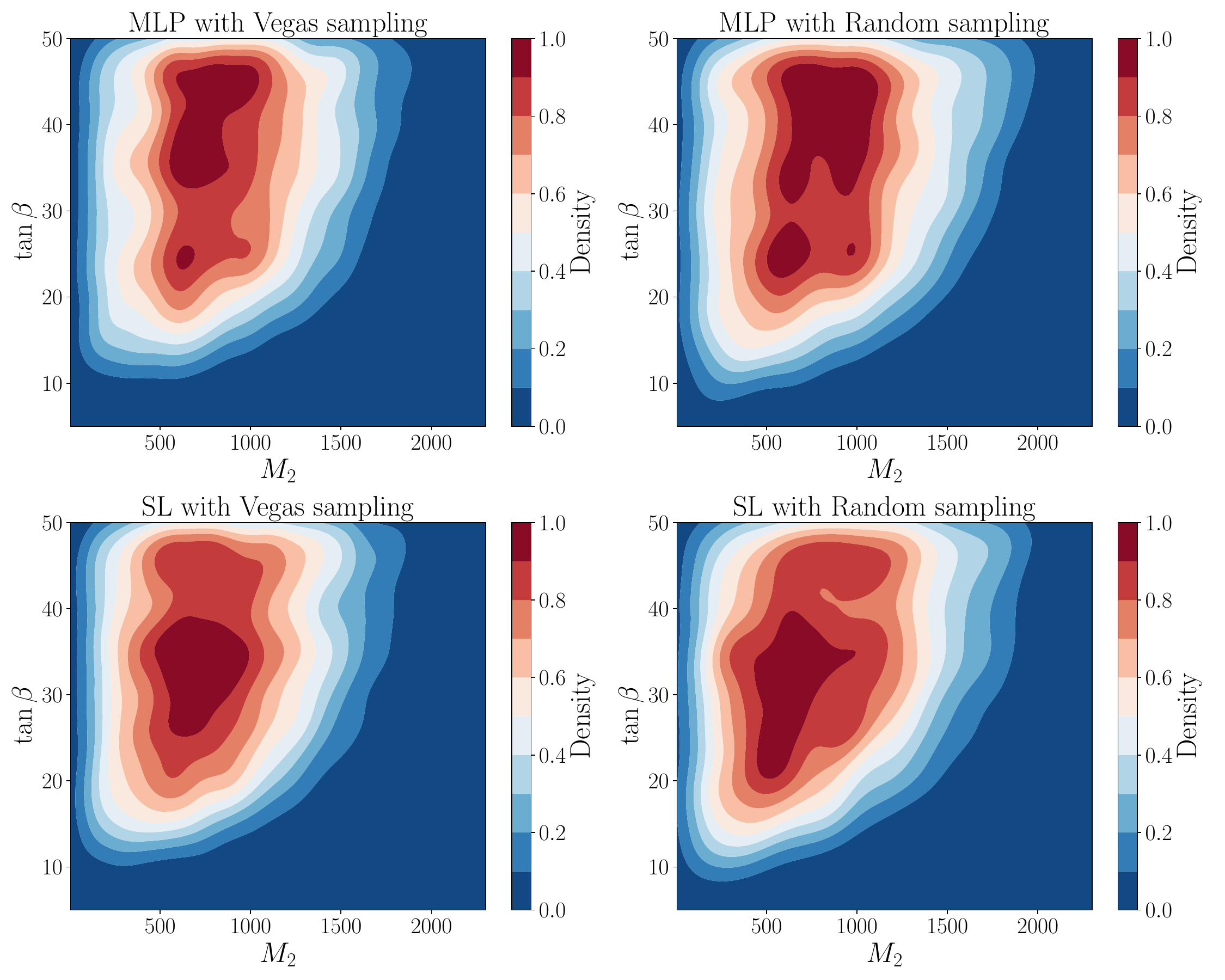}
    \includegraphics[width=0.5\linewidth]{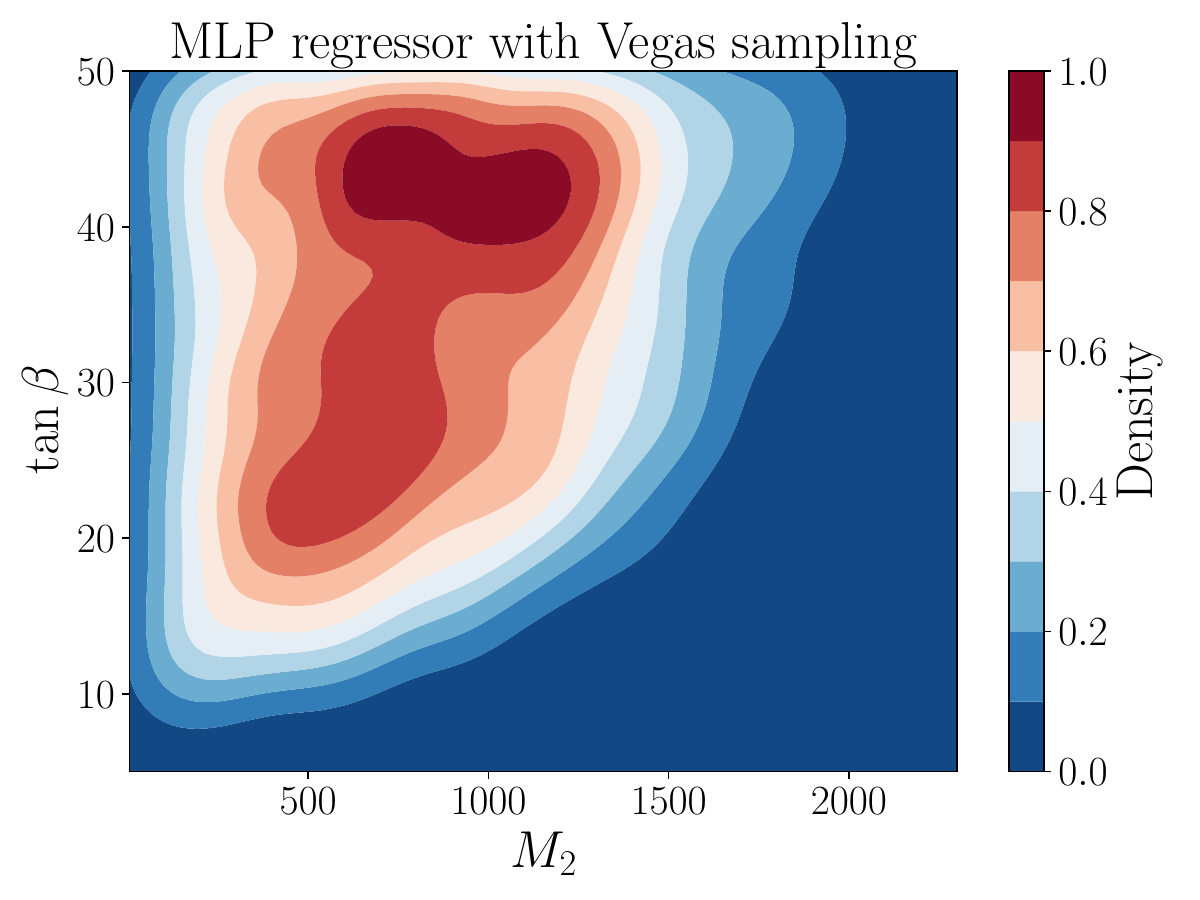}
    \includegraphics[width=0.5\linewidth]{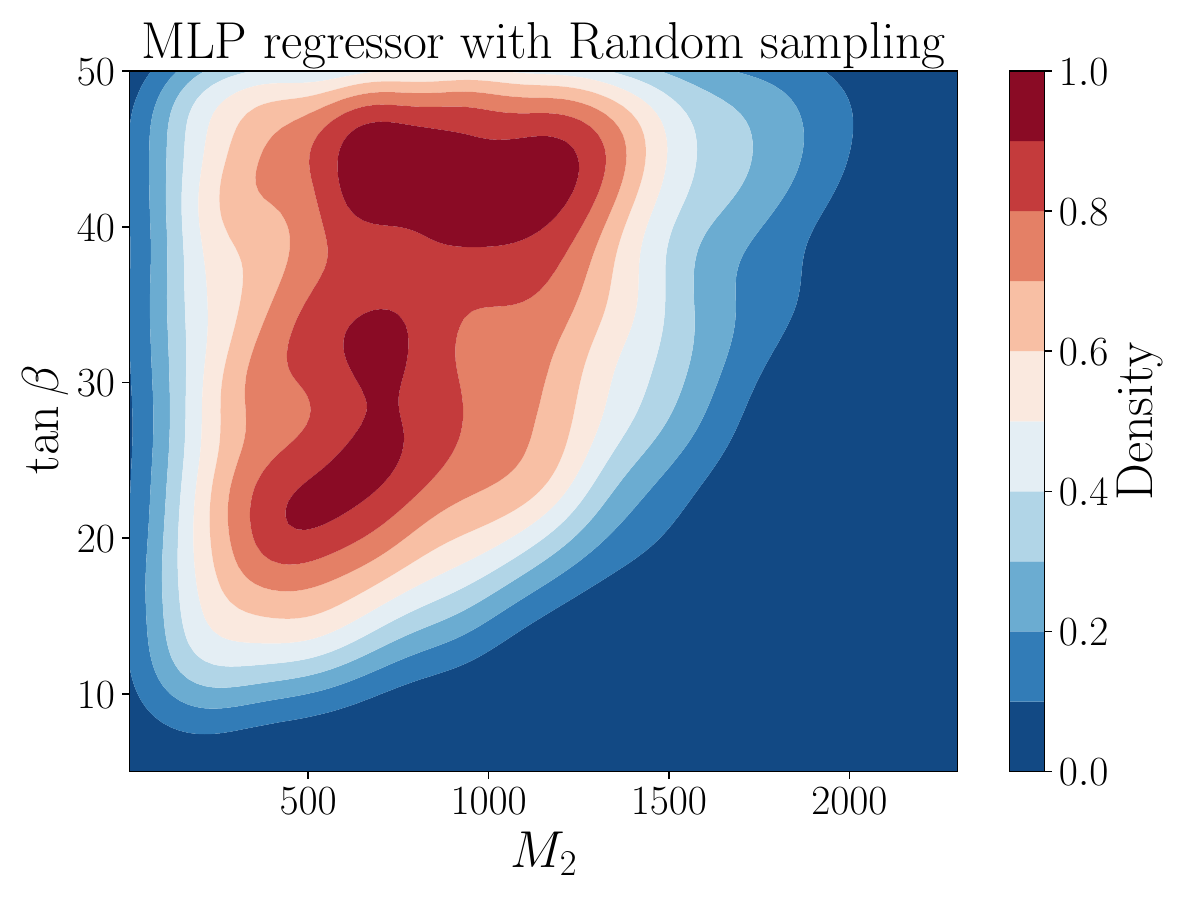}
    \caption{The distribution of the collected valid points in the $(M_2 \text{-} \tan\beta)$ plane highlights the coverage of each scanning method over the target space range. Contour plots for DL with VEGAS sampling are shown in the left panels, while those for DL with random sampling are displayed in the right panels. The color bar indicates the density of the collected valid points in the $(M_2 \text{-} \tan\beta)$ plane. }
    \label{fig:coverage}
\end{figure}

To ensure a fair comparison, all methods are configured with identical hyperparameters when possible. Each method is tasked with collecting 10\,000 valid points, and their performance is evaluated based on the speed of convergence to the target region. The number of iterations required by each method to achieve this goal is presented in Figure~\ref{fig:performance}.

The SL  method combined with VEGAS achieves the best performance, requiring only 53 iterations to accumulate the desired number of valid points. In comparison, the MLP (Multi-Layer Perceptron) with VEGAS takes 57 iterations to achieve the same result. Although SL is generally more effective in high-dimensional spaces, the performance difference in the current analysis is minimal. This is because the scan involves only 5 dimensions and uses a simple target condition, where $m_{h_{\text{SM}}}$ is constrained between 124 and 126 GeV. We anticipate a more distinct performance between the two methods in higher-dimensional scans.

In contrast, when random sampling is employed, SL and MLP require 214 and 316 iterations, respectively, to accumulate the required number of valid points. With VEGAS mapping, SL achieves convergence 4 times faster, while MLP converges 5.5 times faster compared to random sampling. Among the methods evaluated, the MLP regressor approaches exhibit the lowest performance, requiring 417 iterations with VEGAS sampling and 1580 iterations with random sampling. This poor performance of the DL regression methods is consistent with the disadvantages discussed in Sec.~\ref{sec:terminology}.

In any parameter sampling method, coverage is crucial because it ensures the entire parameter space is sufficiently explored, which is essential for obtaining a comprehensive and accurate understanding of the system or model being studied. If certain regions of the parameter space are under-sampled or omitted, the results could be biased. Comprehensive coverage guarantees that all relevant areas are considered, leading to more accurate and representative conclusions. Figure~\ref{fig:coverage} illustrates the coverage area of the collected valid points across all methods used in the $(m_2 \text{-} \tan\beta)$ plane. Interestingly, despite large differences in convergence speed, all methods exhibit similar coverage areas. SL with VEGAS sampling demonstrates higher density in the central region around $\tan\beta = 30$ and $m_2 = 500$ GeV. This is because SL networks cluster valid points closer together in the representation space while pushing invalid points apart. Additionally, the VEGAS mapping generates points closer to areas of high importance. These factors combine to create a narrower, denser region for SL with VEGAS. In contrast, all other methods display a broader coverage area.

Finally, we note that all  DL-based methods, whether classifiers or regressors, exhibit superior performance and faster convergence compared to adaptive sampling techniques such as MCMC or MultiNest, as highlighted in Ref.~\cite{Hammad:2022wpq}.
\section{Conclusion}
\label{sec:conclusion}

The constant experimental efforts keep reducing the space for BSM theories
and in some cases, requiring extensions that add new parameters.
These means that not only we have a reduction in the parameter space that is allowed
but we also have more complicated models to explore.
To this add the increase in experimental tests that cannot be neglected anymore
due to a considerable reduction in their error bars.
All these conditions come together whenever we want to calculate the observables
of a proposed BSM extension and determine the allowed parameter space.
With time, one can only expected an steady increase in the difficulty in testing new
anomalies that these extensions attempt to explain.

Here we presented a new tool to efficiently perform scans
of multidimensional parameter spaces in BSM theories.
By leveraging the power of DL,
we devised an iterative process to predict points likely relevant
in an study, test those points,
train the network, and go back to prediction again.
To solve the problem of efficient generation of test points
we include a VEGAS map that is trained on the collected points
and avoids the inefficiency related to using rejection sampling
in several dimensions.
The result is a tool that can accelerate sampling of high dimensional regions
even with calculations that can be time consuming and computationally expensive,
by preselecting samples from a much larger sample using DL.
We consider two broad types of study: classification and regression.
On the classification side
we treat problems where one would be interested in knowing
a boundary around a targeted result,
for example, when dealing with confidence boundaries or
rejection conditions.
On the regression side,
we deal with problems where it would be useful to actually know the shape
of a quantity inside the parameter space
or consider some importance when collecting points.
Additionally, we consider the possibility of using different architectures
for the neural network and DL techniques,
such as the simple and well known MLP and similarity learning.

This tool has been developed to integrate some well known tools
while at the same time allow for freedom to implement other user required tools.
While we are releasing a first working version,
we expect more ideas for extensions to appear in the future,
mostly with continued and diverse use.
With this in mind, we have developed this tool as open source and
made it available to the wide community via the PyPI repository.
We are excited about the future of this tool,
as ourselves are already using it in real studies
that will be published in the near future,
and keep thinking on future directions to further improve
and test the boundaries of this technique.


\subsection*{Acknowledgments}
AH is funded by grant number 22H05113, ``Foundation of Machine Learning Physics'', Grant in Aid for Transformative Research Areas and 22K03626, Grant-in-Aid for Scientific Research (C). 
The work of RR is supported by a KIAS Individual Grant (QP094601)
via the Quantum Universe Center at Korea Institute for Advanced Study.

\appendix

\section{Generic sampler and its attributes: code example}
\label{sec:genexample}

In this Appendix we display a detailed example of the \texttt{sampler} class (and its attributes)
from \texttt{DLScanner.gen\_scanner} to sample a user defined \texttt{user\_function}
from a user defined module \texttt{user\_model}.
We tried to showcase the use of the most relevant attributes of the \texttt{sampler} class
described in Sec.~\ref{sec:genericscan}.
This code was used to obtain the results of Sec.~\ref{sec:genericscan}
with \texttt{user\_function} defined as in Eq.~\eqref{eq:3dclass}.

\begin{lstlisting}[language=Python]
import numpy as np
import matplotlib.pyplot as plt
from DLScanner.gen_scanner import sampler
from user_model import user_function


# Small function for 3D plotting
def scatter3d_plot(data, title=None, alpha=1.0, savefile=None):
    plt.figure()
    ax = plt.axes(projection='3d')
    if title is not None:
        ax.set_title(title)
    ax.scatter3D(*data.T, s=1, alpha=alpha)
    if savefile is not None:
        plt.savefig(savefile)
    plt.show()


# %%%%%%%%%%%%%%%%%%%%%%%%%%%%%%%%%
# %% sampler and sampler.advance %%
# %%%%%%%%%%%%%%%%%%%%%%%%%%%%%%%%%
# setup
ndim = 3
limits = [[-10*np.pi, 10*np.pi]]*ndim
hidden_layers = 4
neurons = 100
epochs = 1000
use_vegas_map = True
vegas_frac = 0.5
verbose = 1
K = 10000

# Instantiate sampler and do first training
my_sampler = sampler(
    user_function, ndim, limits=limits, K=K,
    method='Classifier', epochs=epochs,
    verbose=verbose,
    use_vegas_map=use_vegas_map, vegas_frac=0.5
)

steps = 8
my_sampler.advance(steps)

# %%%%%%%%%%%%%%%%%%%%%%%%%%%%%%%%%%%%%%%%%%%%%%%%%%%%
# %% sampler.sample and sampler.sample_list         %%
# %% sampler.sample_out and sampler.sample_out_list %%
# %%%%%%%%%%%%%%%%%%%%%%%%%%%%%%%%%%%%%%%%%%%%%%%%%%%%
samples = my_sampler.samples
samples_out = my_sampler.samples_out.flatten()
# Discard some burn-in steps from the beginning
n_dis = 3  # Number of initial samples to discard
samples_bi = np.concatenate(
    my_sampler.samples_list[n_dis:]  # Apply some burn-in
)
samples_bi_out = np.concatenate(
    my_sampler.samples_out_list[n_dis:]
).flatten()


scatter3d_plot(
    samples, title="Accumulated samples", alpha=0.2,
    savefile="in_out_samples.png"
)

scatter3d_plot(
    samples[samples_out == 1], title="In-target samples", alpha=0.5,
    savefile="in_samples.png"
)

scatter3d_plot(
    samples_bi[samples_bi_out == 1], title="In-target samples (burn-in)", alpha=0.5,
    savefile="in_samples_bi.png"
)

# %%%%%%%%%%%%%%%%%%%%%%%
# %% sampler.histories %%
# %%%%%%%%%%%%%%%%%%%%%%%
# Show improvement of loss function with iterative trainings
for j in range(len(my_sampler.histories)):
    plt.plot(
        my_sampler.histories[j].history['loss']
    )
plt.yscale('log')
plt.savefig("histories.png")

# %%%%%%%%%%%%%%%%%%%%%%%%%%%
# %% sampler.vegas_map_gen %%
# %% sampler.model         %%
# %%%%%%%%%%%%%%%%%%%%%%%%%%%
veg_sample, _ = my_sampler.vegas_map_gen(K*100)
pred_labels = my_sampler.model(veg_sample).numpy().flatten()
true_labels = user_function(veg_sample)

scatter3d_plot(
    veg_sample, title="VEGAS map samples", alpha=0.1,
    savefile="vegas_map_gen.png"
)

scatter3d_plot(
    veg_sample[pred_labels > 0.5][:K], title=r"DL selected samples", alpha=0.2,
    savefile="vegas_model_pred.png"
)

sam_selK = veg_sample[pred_labels > 0.5][:K]
true_lab_K = user_function(sam_selK)

scatter3d_plot(
    sam_selK[true_lab_K > 0.5], title="In-target collected samples", alpha=0.2,
    savefile="vegas_pred_intarget.png"
)
\end{lstlisting}

\bibliographystyle{JHEP}
\bibliography{biblo}
\end{document}